\documentclass{aa}  
\usepackage{natbib}
\usepackage{graphicx}
\usepackage{appendix}
\usepackage{txfonts}
\usepackage{dcolumn}
\begin{document}

   \title{The distribution of stars around the Milky Way's 
   central  black hole: I. Deep star counts}

   %\subtitle{I. Overviewing the $\kappa$-mechanism}

   \author{E. Gallego-Cano
          \inst{1}
          \and
          %C. Ptolemy\inst{2}\fnmsep\thanks{Just to show the usage
          %of the elements in the author field}
          R. Sch\"odel
          \inst{1}
          \and
          H. Dong
          \inst{1}
          \and
          F. Nogueras-Lara           
          \inst{1}
          \and
          A. T. Gallego-Calvente
          \inst{1}
          \and
          P. Amaro-Seoane
          \inst{2}
          \and
          H. Baumgardt
          \inst{3}
          }

   \institute{
    Instituto de Astrof\'isica de Andaluc\'ia (CSIC),
     Glorieta de la Astronom\'ia s/n, 18008 Granada, Spain
              \email{lgc@iaa.es}
         \and
       Institut de Ci{\`e}ncies de l'Espai (CSIC-IEEC) at Campus UAB, Carrer de Can Magrans s/n 08193 Barcelona, Spain\\
      Institute of Applied Mathematics, Academy of Mathematics and Systems Science, Chinese Academy of Sciences, Beijing 100190, China\\
   Kavli Institute for Astronomy and Astrophysics, Beijing 100871, China\\
 Zentrum f{\"u}r Astronomie und Astrophysik, TU Berlin, Hardenbergstra{\ss}e 36, 10623 Berlin, Germany
         \and
      School of Mathematics and Physics, University of Queensland
      St. Lucia, QLD 4068,Australia
             }

   \date{Received; accepted }

% \abstract{}{}{}{}{} 
% 5 {} token are mandatory
 
  \abstract
  {The existence of dynamically relaxed stellar density cusp in dense
    clusters around massive black holes is a long-standing prediction
    of stellar dynamics, but it has so far escaped unambiguous
    observational confirmation.}
   { In this paper we aim to revisit the problem
    of inferring the innermost structure of the Milky Way's nuclear
    star cluster via star counts, to clarify whether it displays
    a core or a cusp around the central black hole.}
   {We used judiciously selected adaptive optics assisted high angular resolution images
     obtained with the NACO instrument at the ESO VLT. Through image
     stacking and improved PSF fitting we pushed the completeness limit
     about one magnitude deeper than in previous, comparable
     work. Crowding and extinction corrections were derived and applied
   to the surface density estimates. Known young, and therefore
   dynamically not relaxed stars, are excluded from the
   analysis. Contrary to previous work, we analyse the stellar density in
   well-defined magnitude ranges in order to be able to constrain stellar masses
   and ages.}
 {We focus on giant stars, with observed magnitudes $K=12.5-16$,
     and on stars with observed magnitudes $K\approx18$, which may
     have similar mean ages and masses than the former. The giants
     display a core-like surface density profile within a projected
     radius $R\leq0.3$\,pc of the central black hole, in agreement
     with previous studies, but their 3D density distribution is not
     inconsistent with a shallow cusp if we take into account the
     extent of the entire cluster, beyond the radius of influence of
     the central black hole. The surface density of the fainter
     stars can be described well by a single power-law at $R<2$\,pc.
     The cusp-like profile of the faint stars persists even if we take
     into account the possible contamination of stars in this
     brightness range by young pre-main sequence stars. The data are
     inconsistent with a core-profile for the faint stars. Finally, we
     show that a 3D {\it Nuker} law provides a good description of the
     cluster structure.}
 { We conclude that the observed density of the faintest stars
   detectable with reasonable completeness at the Galactic Centre, is
   consistent with the existence of a stellar cusp around the Milky
   Way's central black hole, Sagittarius\,A*. This cusp is well
     developed inside the influence radius of Sagittarius\,A* and
     can be described by a single three-dimensional power-law with an
     exponent  $\gamma=1.43\pm0.02\pm0.1_{sys}$.  This
     corroborates existing conclusions from Nbody simulations
     performed in a companion paper. An important caveat is that the
     faint stars analysed here may be contaminated significantly by
     dynamically unrelaxed stars that formed about 100\,Myr ago. The
   apparent lack of giants at projected distances of
   $R\lesssim0.3\,pc$ ($R\lesssim8"$) of the massive black hole may
   indicate that some mechanism may have altered their distribution or
   intrinsic luminosity. We roughly estimate the number of possibly
   missing giants to about 100.} 


   \maketitle
%
%-------------------------------------------------------------------

\section{Introduction}

This is the first one of three papers addressing the distribution of
stars around Sagittarius\,A* (Sgr\,A*), the massive black hole at the
centre of the Milky Way. They are closely related, but focus on
different methods and stellar populations. In this work we use the
method of star counts, while in our other paper (Sch{\"o}del et al.,
referred to as Paper II in the following), we analyse the diffuse
light from the unresolved stellar population. Finally, Baumgardt et
al.\ (Paper\,III) present new Nbody simulations that confirm the
agreement between modelling and observations.

The distribution of stars around Sgr\,A* is of great astrophysical
interest because it is the only massive black hole where we can test
observationally the existence of a stellar cusp. Such a stellar cusp
is a prediction of stellar dynamics for the case of a dynamically
relaxed cluster and has been derived and studied by analytical, Monte
Carlo and N-body methods
\citep[e.g.][]{Bahcall:1976vn,Lightman:1977ly,Murphy:1991zr,Baumgardt:2004fk,Amaro-Seoane:2004kx,Alexander:2009gd,Preto:2010kx}.
These consistent theoretical results have, however, not yet been
confirmed observationally. There exist currently only measurements of
about two dozen systems, where we can actually resolve the radius of
influence of the central black hole with several resolution elements
\citep[see, e.g. Table\,1 in][]{Gultekin:2009fk}. The great distance
of most extragalactic systems means that we can only study the light
density of hundreds to thousands, or even millions, of stars per
resolution element. Since most nuclear star clusters are entities with
complex stellar populations, many of which show signs of recent star
formation \citep[see review by ][]{Boker:2010ys}, measured light
densities may frequently be dominated by a small number of bright
stars, which are generally too young to be dynamically relaxed. This
can lead to ambiguous or erroneous results.

On the other hand, with its distance of only about 8\,kpc from Earth,
the Galactic Centre (GC) allows us to resolve the stars
observationally on scales of about 2\,milli-parsecs (mpc), assuming
diffraction limited observations at about $2\,\mu$m at an 8-10m
telescope. At the centre of the Milky Way, a
$4\times10^{6}$\,M$_{\odot}$
\citep[e.g.][]{Boehle:2016zr,Gillessen:2016uq} massive black hole,
Sagittarius\,A* (Sgr\,A*), lies embedded in a
$\sim$$2.5\times10^{7}$\,M$_{\odot}$
nuclear star cluster \citep{Schodel:2014fk,Schodel:2014bn,Feldmeier:2014kx,Chatzopoulos:2015yu,Feldmeier-Krause:2017rt}.
Therefore, the GC appears to be, in principle, the ideal test case for
the existence of stellar cusps.

Surprisingly, no unambiguous observational evidence for the existence
(or not) of a stellar cusp around Sgr\,A* has been presented so
far. While the first high angular resolution observations at an 8m
telescope appeared to indicate the existence of a stellar cusp
\citep{Genzel:2003it,Schodel:2007tw}, it was later realised that the
star counts within about 0.5\,pc of Sgr\,A* were contaminated by a
significant number of young, and therefore dynamically unrelaxed,
stars. When omitting the young stars, the projected stellar density of
giants appears almost flat, core-like, within a few $0.1$\,pc
of Sgr\,A*
\citep{Buchholz:2009fk,Do:2009tg,Bartko:2010fk}. Observations of the
stellar surface brightness from old stars also appeared to indicate a
possibly core-like structure \citep{Fritz:2016fj}. These findings led
to the missing cusp problem and have given rise to a large
number of theoretical papers, trying to explain its absence. The
hypotheses reach from a very long relaxation time
\citep[][]{Merritt:2010ve}, over the destruction of the envelopes of
giants -- thus rendering them invisible -- via stellar collisions,
which cannot fully explain the observations \citep{Dale:2009ca} or a
fragmented gaseous disc, which
can \citep{Amaro-Seoane:2014fk}. Other explanations involve the
apparent stellar structure arising from subsequent epochs of star
formation and/or the accumulation of stellar remnants near Sgr\,A*
\citep[e.g.][]{Lockmann:2010fk,Aharon:2015uq}.

When evaluating the observational evidence, it is, however, of utmost
importance to be aware of its limitations. Firstly, due to the extreme
interstellar extinction towards the Galactic Centre
\citep[e.g.][]{Nishiyama:2009oj,Schodel:2010fk,Fritz:2011fk},
observations need to be performed in the near-infrared (NIR). A second
requirement for observing the Galactic Centre (GC) is to use an
angular resolution as high as possible to overcome the extreme source
crowding. Here we use the adaptive optics (AO) assisted NIR camera
NACO installed at the ESO VLT. Imaging data at $H$ and $K_{S}$ are
used to be able to estimate extinction and -- by the same means -- to
exclude foreground stars.

Because of these observational difficulties, our knowledge about the
stellar population at the GC is limited to the brightest few percent
of stars: A few million-year-old hot post main sequence (MS) giants
and MS O/B stars (the latter being already at the faint limit of
spectroscopic capabilities), on the one hand, and, on the other hand,
giants with luminosities equal to or higher than RC
stars. In fact, the typical spectroscopic completeness reaches only
about $K_{S}=15.5$ stars and thus only half the RC
\citep[see][]{Do:2009tg,Bartko:2010fk,Do:2015ve,Stostad:2015qf,Feldmeier-Krause:2015}. Studies of the
stellar surface density at the GC have so far been dominated by RC
stars and brighter giants. The only recent study that focussed on the light from stars fainter
than this limit did indeed find a cusp-like structure within
$5"/0.2$\,pc of Sgr\,A* \citep{Yusef-Zadeh:2012pd}.  Hence, we have
only observed the tip of the iceberg, which may not be representative
for the overall, underlying structure.

This work continues similar efforts carried out by
\citet{Genzel:2003it} and \citet{Schodel:2007tw}. As experiments are
repeated, both their accuracy and precision tend to increase because
of factors such as improvements in observational techniques,
increasing know-how on data reduction and analysis, progress in theory
and interpretation, and an increasing amount of data. The novel
aspects of this work are, in particular, the stacking of high quality
images with a large field-of-view (FOV) to reach fainter completeness
limits in the most crowded regions near Sgr\,A*, an extension of high
angular resolution data to a larger field of about $1.5'\times1.5'$,
improvements in data reduction (rebinning of the images, removal of
systematic noise from detector electronics), and analysis (improved
PSF fitting with use of noise maps and a spatially variable
PSF, explicit consideration of systematic uncertainties caused
  by choice of parameters in the PSF fitting code). Finally, more explicitly than in previous work -- and because
our deeper data allow us to do so -- we focus on clearly delimited
stellar brightness ranges in order to minimise the mixing of different
stellar populations. We add new data on the stellar distribution at
projected radii $R\gtrsim2$\,pc from the literature
\citep{Schodel:2014fk,Fritz:2016fj} to facilitate the interpretation
of the data and the derivation of the 3D density structure of the
stars near the massive black hole.

%--------------------------------------------------------------------
\section{Data reduction and analysis}

\subsection{Basic reduction} 
We use $H$ and $Ks$-band data obtained with the S27 camera
($0.027"$ pixel scale) of NACO/VLT. The AO was locked on the NIR bright
supergiant GCIRS\,7 that is located about $5.5"$ north of Sgr\,A*. The
data used are summarised in Table\,\ref{Tab:Obs}. All data were
acquired with a similar four point dither pattern, roughly centered on Sgr\,A*,
with the exception of the data from 11 May 2011, which covered a
shallow, but wider mosaic with a $4\times4$ dither pattern,
centered on Sgr\,A*. Preliminary data reduction was standard, with sky
subtraction, bad pixel removal and flat fielding. Subsequently, a
simple shift-and-add (SSA) procedure was applied to obtain final images. A
quadratic interpolation with a rebinning factor of two was used because
tests showed that this improved the photometry and reduced residuals
in PSF fitting, in particular since the S27 pixel scale barely samples
the angular resolution, which was roughly $0.06"$ FWHM for all
images. Along with the mean SSA images we also created noise maps that
contain the error of the mean of each pixel in the SSA images.

\begin{table}
\centering
\caption{Details of the imaging observations used in this
  work.}
\label{Tab:Obs} 
\begin{tabular}{llllll}
\hline
\hline
Date$^{\mathrm{a}}$ & $\lambda_{\rm central}$ & $\Delta\lambda$ & N$^{\mathrm{b}}$ & NDIT$^{\mathrm{c}}$ & DIT$^{\mathrm{d}}$\\
 &  [$\mu$m]  &   [$\mu$m] &  & & [s] \\
\hline
09 May 2010 & 1.66 & 0.33 &  4 & 64 & 2 \\
17 May 2011 & 2.18 & 0.35 & 4 & 9 & 2  \\
09 Aug 2012 & 2.18 & 0.35 & 8 & 60 & 1  \\
11 Sep 2012 & 2.18 & 0.35 & 8 & 60 & 1  \\
12 Sep 2012 & 2.18 & 0.35 & 8 & 60 & 1  \\
\hline
\end{tabular}
\begin{list}{}{}
\item[$^{\mathrm{a}}$] UTC date of beginning of night.
\item[$^{\mathrm{b}}$] Number of (dithered) exposures
\item[$^{\mathrm{c}}$] Number of integrations that were averaged on-line by the read-out
  electronics
\item[$^{\mathrm{d}}$] Detector integration time. The total integration time of each observation amounts to N$\times$NDIT$\times$DIT.
\end{list}
 \end{table}

\begin{figure*}[!htb]
\includegraphics[width=\textwidth]{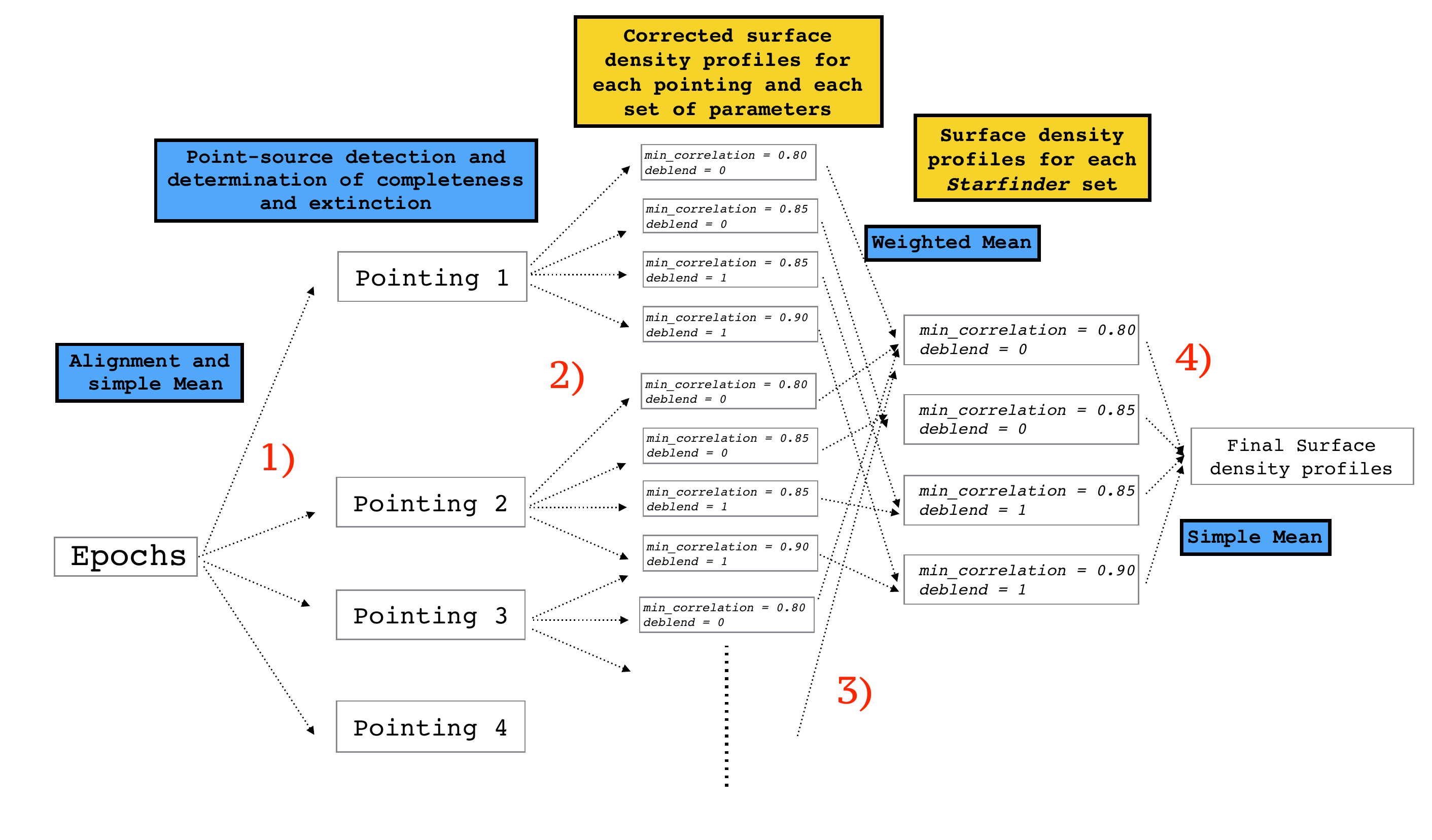}
\caption{\label{Fig:Scheme} Scheme of the procedure to compute the
  surface density profiles in the deep mosaic. The four pointings are
  analysed independently to avoid distortion problems. In step (1) the
  images from the different epochs for each of the four pointings are
  combined. In step (2), we analysed the photometry for the four
  pointings with four different values of the {\it Starfinder}
  parameter, determined the source detection completeness, computed
  the extinction map. Tables with the photometric and
    astrometric parameters for the point source detections, for each
    pointing and {\it StarFinder} parameter set are available at the
    CDS.} Finally,we computed the extinction- and
  completeness-corrected surface density profiles. In the step (3), we
  computed the surface density profile for all {\it Starfinder}
  parameter sets by combining the results from the four pointings. In
  step (4), the surface density profiles obtained for each {\it
    Starfinder} set are mean-combined.
\end{figure*}

\subsection{Pattern removal} 
The images from the individual epochs contained horizontal stripe
patterns from the detector electronics.  These horizontal stripes can
be detrimental for source detection because they may either mask faint
sources or be deblended into rows of stars by the PSF fitting
program. It is therefore important to remove them. We proceeded as
follows: We used the {\it StarFinder} program to detect and subtract
robustly detected point sources from each image (conservative settings
of the {\it StarFinder} parameters: {\it min\_correlation$=0.85$} and
{\it deblend$ = 0$}) and to fit the diffuse emission (from unresolved
stars or dust and gas in the interstellar medium). The latter was
fitted with an angular resolution of about $0.25"$, a non-critical
value that just needs to be large enough to remove the variable
background due to unresolved stellar emission and small enough to
roughly correspond to the size of diffuse, unresolved structures in
the mini-spiral (see Paper\,II). While fitting of the diffuse
background is important in this procedure, the exact choice of its
variability scale is not. It can easily be chosen to be a factor 2
larger or smaller.

\begin{figure}[!htb]
\includegraphics[width=\columnwidth]{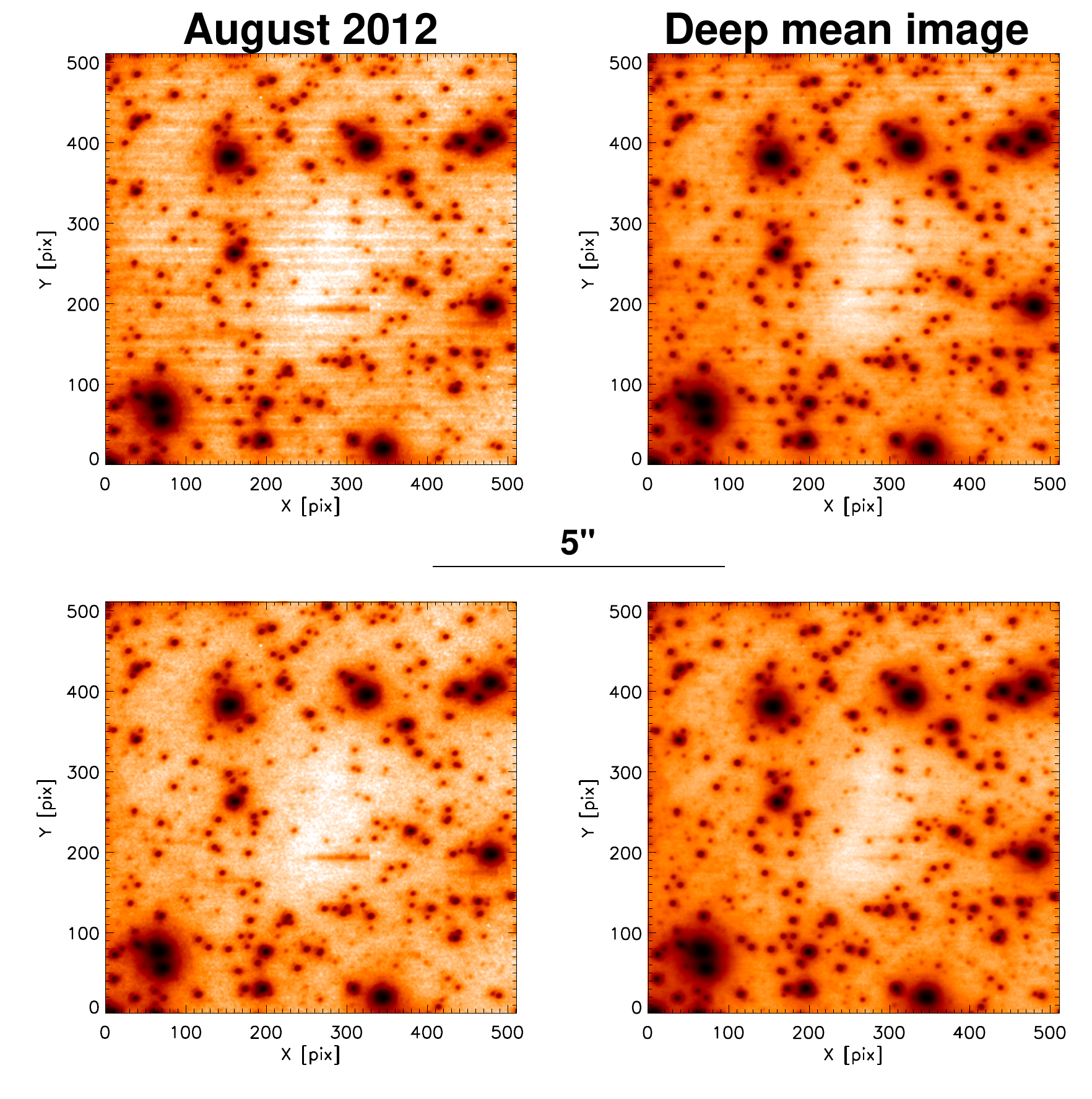}
\caption{\label{Fig:stripes} Cleaning of horizontal stripes
  (systematic readout noise). Upper left: Detail of August 2012
  $Ks$-band image. Lower left: As upper left, but cleaned. Upper
  right: Detail of deep, mean combined $Ks$-band image when the input
  images have not been cleaned.  Lower right:  Detail of deep, mean
  combined $Ks$-band image after cleaning of the input
  images. The displayed field is located about $12.0"$ and $1.7"$
north of Sgr\,A*. The colour scale is logarithmic and identical for
all images.}
\end{figure}

The resulting residual images, i.e.\ image minus diffuse emission
minus point sources, were then dominated by small-scale (on the order
a few pixels width) random and systematic noise. We determined the
pattern of horizontal stripes induced by the electronics through
median smoothing each row of pixels with a median box width of about
$2.7"$, corresponding to 200 pixels (in the rebinned images). This
pattern was then subtracted from the SSA images. We could thus remove
most of the systematic noise without introducing any significant bias
on the point sources or on the diffuse emission because most stars had
already been subtracted and because the median smoothing box was a
factor of a few to ten larger than the scales of the diffuse emission,
of the size of PSF residuals, or of faint, unresolved sources.
Finally, after having cleaned the images of each epoch, they were
combined to a deep mean image (see next section). This last step
further reduced any remaining systematics. To be conservative, we used
the noise maps derived from the uncleaned
images. Fig.\,\ref{Fig:stripes} shows details of $K_{s}$-images to
illustrate the effect of the systematic readout noise and the
improvement after removing it.  Figure\,\ref{Fig:Scheme} shows an
  outline of the procedure that we followed after this basic
  reduction. 

\subsection{Alignment and stacking} 
We treated each of the four pointings towards Sgr\,A* independently to
avoid problems arising from camera distortions near the edges of the
NACO S27 camera's field
\citep[see][]{Trippe:2008it,Schodel:2009zr}. The final images from all
epochs were aligned with the one from 09 Aug 2012 via a polynomial fit
of degree one (IDL POLYWARP and POLY\_2D procedures). The parameters
of the latter were determined via an iterative fit using lists of
detected stars in the image. The images were combined in a simple mean
and the corresponding noise maps were quadratically combined 
  (step\,1 in Fig.\,\ref{Fig:Scheme}). 

A possible concern in this stacking procedure is the use of different
observing epochs because of the large proper motions of the stars in
the GC. At the distance of the GC \citep[here assumed as 8\,kpc, see,
e.g.][]{Genzel:2010fk,Meyer:2012fk}, a velocity of about
40\,km\,s$^{-1}$ on the plane of the sky corresponds to a proper
motion of one milli-arcsecond per year. A displacement by one pixel of
the NACO S27 camera per year therefore corresponds to a velocity
$>1000\,$km\,s$^{-1}$. Since we are not interested in high precision
astrometry or photometry, this effect is therefore negligible for our
data, except possibly a small number of very fast moving stars within
$\sim$$0.1"$
of Sgr\,A*, which is not relevant to the problem and scales addressed
in this paper.

\subsection{Source detection \label{sec:detection}} 
Point source extraction was carried out with the PSF fitting program
{\it StarFinder} \citep{Diolaiti:2000qo}. Since the images cover areas
similar to or larger than the isoplanatic angle at $Ks$-band,
care was taken to deal with the spatial variability of
the PSF. We parted the images into sub-fields of approximately
$10.5"\times10.5"$ size. Subsequently, ten of the brightest, most
isolated stars in each sub-field were used for an iterative extraction
of a local PSF \cite[similar to what was done
in][]{Schodel:2010fk}. Because of variable extinction and source
density, not all sub-fields contained PSF reference stars of similar
brightness, which would lead to a systematic change of the zero point
across the field. Also, not taking into account the extended seeing
halo from the light that could not be corrected by the AO, can lead to
an enhanced detection of spurious faint stars near bright stars. As
remarked by \citet{Schodel:2010hc}, the seeing halo is affected in a
rather minor way by anisoplanatic effects. We therefore used the
brightest star in the field, GCIRS\,7, to estimate the seeing
halo. The local PSFs were masked beyond radii of about $0.3"$, up to
which they could be reliably determined. Then they were matched to the
seeing halo (using a least-squares fit to determine flux offsets and
normalisation factors). Thus, we could create local PSFs that avoided
large systematic photometric effects across the field. Some tests
\citep[similar to what was done in][]{Schodel:2010hc} showed that we
could constrain the systematic photometric effects from the
variability of the PSF to a few percent across the field.

In PSF fitting we have to walk a thin line between achieving an
almost complete detection of sources while, at the same time, avoiding
to pick up spurious ones, which can arise, in particular, close to
bright stars or due to systematic effects from the detector
electronics. We visually verified that taking the PSF seeing halos into
account, along with the use of our SSA noise maps, effectively
suppressed the detection of spurious sources near bright stars
\citep[see also][]{Schodel:2013fk}. The PSF halos include effects such
as diffraction spikes and static speckles. Our empirical noise maps
seemed to deal well with suppressing the detection of spurious sources
near bright stars.

\begin{figure*}[!htb]
\includegraphics[width=\textwidth]{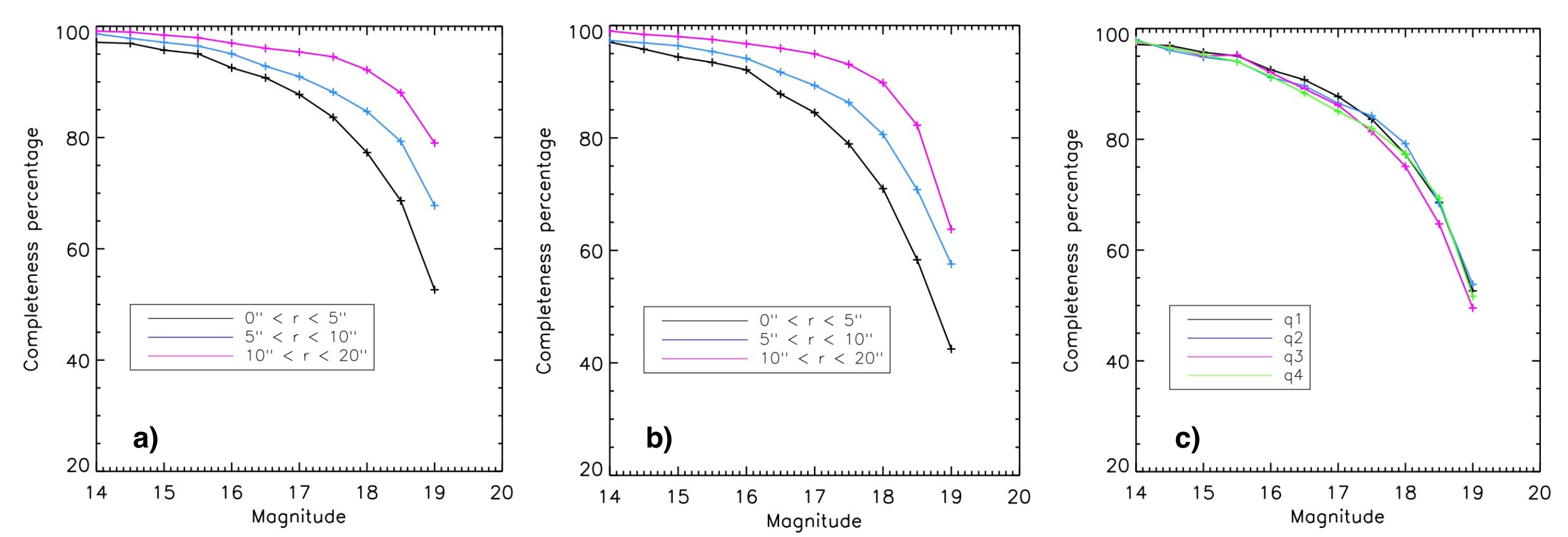}
\caption{\label{Fig:completeness} Completeness of the star counts in
  the deep NACO $K_{s}$ images. a) Completeness in pointing 1 for
  different projected distance ranges from Sgr\,A*, for {\it
    min\_correlation$=0.80$} and {\it deblend$ = 0$}.  b) As in a),
  but for {\it min\_correlation$=0.90$} and {\it deblend$ = 1$}. c)
  Completeness for all four pointings and within $5"$ of Sgr\,A*, for
  {\it min\_correlation$=0.80$} and {\it deblend$ = 0$}. The
  corresponding plots for other used combinations of {\it min\_correlation
    and deblend} look very similar.}
\end{figure*}

Finally, since there can be no absolute certainty in the reliability
of source detection, we also repeatedly analysed the images with
different values of the {\it StarFinder} parameters that dominate the
probability of source detection (for a fixed detection threshold,
which was chosen as $3\sigma$ in all cases). These parameters are {\it
  min\_correlation} and {\it deblend}. For the minimum correlation
value we chose $0.8-0.9$, always more conservative than the standard
value of $0.7$, the default value of {\it StarFinder}. The key word
{\it deblend} can be set to deblend close sources. While deblending can be
very useful, it can lead also to the detection of a significant number
of spurious sources in a crowded field. We included measurements with
and without setting this keyword. We used the following four
  combinations of {\it min\_correlation} and {\it deblend}: $[0.80,0],
  [0.85,0], [0.85,1]$, and $[0.90,1]$  (step 2 in Fig. \,\ref{Fig:Scheme}).

\subsection{Photometric calibration and source selection}
 Finally, the photometry was calibrated with the stars IRS\,16C,
 IRS\,16NW, and IRS\,33N \citep[apparent magnitudes
 $K_{s}=9.93,10.14,11.20$ and
 $H=11.90,12.03,13.24$, see][]{Schodel:2010fk}. The uncertainty of the zero
 points was a few percent. We note that for the purposes in this paper
 we do not require any high accuracy and high precision photometry and
 astrometry.

 Almost all stars in the field have intrinsic colours
 $-0.1\leq H-K \leq 0.3$ \citep[see,
 e.g.][]{Do:2009tg,Schodel:2010fk}. The mean colour due to reddening
 is $H-K\approx2.1$. We excluded all stars with $H-K<1.5$ as
 foreground stars. We also excluded spectroscopically identified young
 stars from our final star list
 \citep{Do:2009tg,Bartko:2010fk}. Subsequently, we created an
 extinction map, by using the 20 stars nearest to each point. The
 resulting map is similar, to within the uncertainties, to the one
 presented in \citet{Schodel:2010fk}.

\subsection{Crowding and completeness \label{sec:completeness}} 

We determined the source detection completeness in the $K_{s}$-images
through the technique of inserting and recovering artificial stars
for  each of the four pointings (the second step in Fig.\,\ref{Fig:Scheme}) . We
used a magnitude step of $0.5$\,mag and inserted the stars on a
$0.5"\times0.5"$ grid. With this relatively wide spacing we avoided
artificially increasing the crowding. The grid was shifted several
times to finally probe completeness on a dense $0.1"\times0.1"$ grid
\citep[as done by ][]{Schodel:2007tw}. We used the respective local
PSFs (see above). Subsequently, PSF fitting was carried out with {\it
  StarFinder} and a source was considered as detected if it was found
within a magnitude range of 0.5\,mag of the input magnitude and within
a distance of $0.054"$ of the input position (corresponding to 2
pixels of the S27 camera or roughly the angular resolution of the
data). If a real star of a similar magnitude was already present
within this distance to the grid point of an artificial star, then the
artificial star was considered as detected. This latter point is
critical to avoid bias because the relatively high density of
artificially introduced stars would otherwise lead to non-detection of
real sources and thus an over-estimation of incompleteness.

As mentioned above, to estimate the systematic errors induced by
either the non-detection of real sources or the detection of spurious
sources, we repeated the source detection and completeness
determination for the following combinations of the {\it StarFinder}
parameters: {\it min\_correlation$=0.80,0.85,0.85,0.90$} and {\it
  deblend$ = 0, 0, 1, 1$} (each value in the first list corresponds to
the value with the same index in the second list). In
Fig.\,\ref{Fig:completeness}, we show the values of completeness for
two of these cases and for different projected distance ranges from
Sgr\,A*. The differences between the different choices of parameters
are generally small, on the order of a few percentage points, except
for the faintest magnitudes, where the differences are somewhat more
pronounced. Also, we can observe the expected general trend of less
completeness for fainter magnitudes and in the more crowded areas near
Sgr\,A*. Finally, Fig.\,\ref{Fig:completeness} also shows that the
differences of completeness between the four pointings are small. For
all cases, we found that source detection was, at all projected
distances, at least 50\% complete for magnitudes $K_{s}\leq18.5$ (Step
2 in Fig. \,\ref{Fig:Scheme}).

\subsection{Extinction}

\begin{figure}[!htb]
\includegraphics[width=\columnwidth]{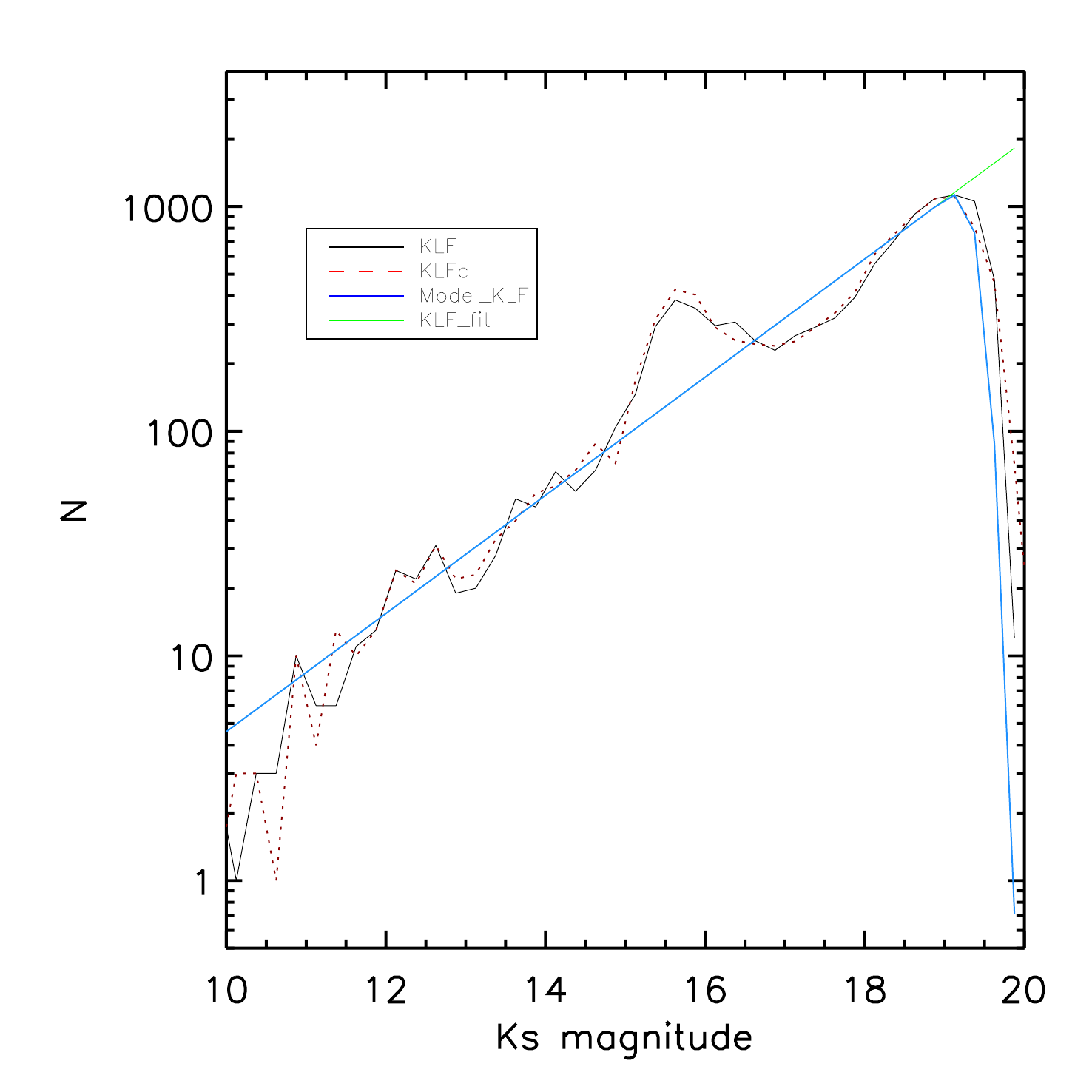}
\caption{\label{Fig:ModeledKLF} The Ks LF in pointing 1 for {\it
    min\_correlation$=0.8$} and {\it deblend$=0$} if shown as a black
  line. The dotted red line is the $K_{s}$ LF after correcting the 
  magnitude of each star for differential extinction. The green lines shows
  a power law fit to the bright stars $11<K_{s}<14.5$ with power law
  index of $0.26\pm0.02$. The blue line shows the effect of the
  completeness function according to equation $(2)$ in \citet{Chatzopoulos:2015uq}.}
\end{figure}
 
We used the $H-K_{s}$ photometry and the intrinsic small colours of
stars at these bands to create an extinction map for each of the four
  pointings (the second step in Fig.\,\ref{Fig:Scheme}),
with the same method as applied in \citet{Schodel:2010fk}. We do not
consider stars with $H-K_{s}<1.5$ because we consider them as
foreground stars. Neither do we consider stars with $H-K_{s}>3.0$
because they may either be background stars or intrinsically reddened
objects \citep[in any case, their number is very small, see Fig\,4
in][]{Schodel:2010fk}. Median stellar colours were obtained from the
individual colours of the 20 nearest stars at each position and the
extinction was then calculated as in \citet{Schodel:2010fk}, assuming
$A_{K_{s}}\propto\lambda^{-2.2}$.
%We are
%mainly interested in the extinction variation within the Galactic
%center, therefore we computed a relative extinction map for each
%pointing.
% We have to consider that the effect of the extinction is,on
%the one hand, to weaken the star magnitudes and on the other hand it
%does that many stars remain hidden. Therefore, we address two
%corrections. 

On the one hand, the extinction map was used to correct the individual
stellar magnitudes for differential extinction. On the other hand, we
applied the methodology of \citet{Chatzopoulos:2015uq} to compute the
stellar detection completeness variation caused by variable
extinction: We modelling the luminosity function (LF) by taking the
product of a power-law stellar LF and an error function that
represents the completeness function, as in expression $(2)$ in
\citet{Chatzopoulos:2015uq}. The approximation of the LF with a
power-law -- which ignores the presence of the RC bump --
does not introduce any significant error because our data are
sensitive enough to reach well below the RC bump over the entire field
and because including the RC bump would only have a minor effect as
shown by \citet{Chatzopoulos:2015uq}. First, we measured the observed
KLF for each pointing and each {\it StarFinder} parameter set (excluding a
radius of about 5'' around Sgr\,A*, where the KLF is more incomplete
because of crowding). We computed the power law for each
case. Finally, we used these power-laws, combined it with the measured
local extinction and computed the corresponding local correction
factors according to equation $(5)$ in \citet{Chatzopoulos:2015uq},
but using the approximation of a single extinction screen, i.e.\ no
variability of $A_{K_{s}}$ along the line-of-sight. Since we
approximate the LF with a power law, the equation takes on the form
{$p = L(-\Delta A_{k})=10^{-\gamma*\Delta A_{k}}$} \citep[see][]{Chatzopoulos:2015uq}, where $p$ is the
reduction factor for the number of locally detected stars,
$\Delta A_{k}$ is the difference between the local extinction and the
mean extinction over the field, and {$\gamma$} is the power law index
of the luminosity function. If the local extinction is lower than the
mean extinction, then $p>1$, and if the local extinction is higher
than the mean extinction, then $p<1$.

\begin{figure}[!htb]
\includegraphics[width=\columnwidth]{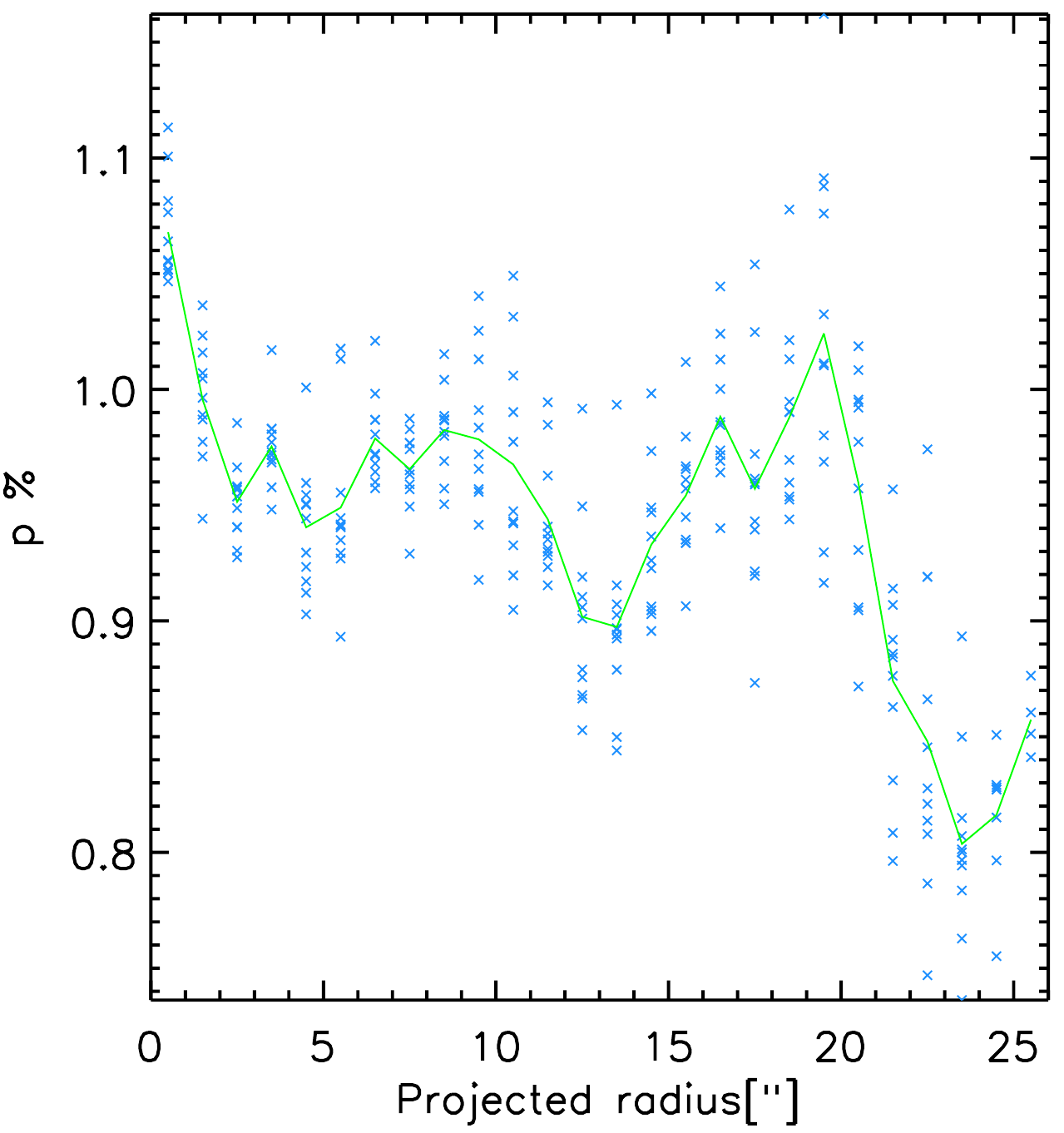}
\caption{\label{Fig:percentage_chatz} Relative detection frequency due
  to extinction versus projected distance to Sgr\,A*, in pointing 1
  for {\it min\_correlation$=0.8$} and {\it deblend$=0$}. The blue
  crosses give the values of $p$ for each magnitude bin. The green
  line represents the mean of the {$p(\%)$} considering detected stars
  at the same distance from Sgr\,A*. We can observe that for close
  distances to Sgr\,A*{$p(\%)$} is higher than for large distances, as
  we expected, because the extinction near Sg\,A* is lower.}
\end{figure}

We apply the correction factor $1/p$ to each detected star. In
Fig.\,\ref{Fig:ModeledKLF} we show the $K_{s}$ LF for pointing 1 , for
{\it min\_correlation$=0.8$} and {\it deblend$=0$}, along with the LF
corrected for differential extinction, a power law fit to the stars
$11<K_{s}<14.5$ ( $\gamma=0.26\pm0.02$,similar to the value obtained
in \cite{Schodel:2010fk}), and the completeness function (blue line),
as defined by \citet{Chatzopoulos:2015uq}. For the latter, we use
$m_{0}=19.4$ and $\sigma=0.2$ because these values approximate our
$K_{s}$ LF well. These values are different from those used in
\citet{Chatzopoulos:2015uq} because our data are significantly deeper
than theirs. In Fig.\,\ref{Fig:percentage_chatz} the percentage
reduction in observed stars versus projected radius is represented for
the detected stars in pointing 1,for {\it min\_correlation$=0.8$} and
{\it deblend$=0$}. One can see that extinction is, on average, higher
at larger distances from Sgr\,A*, but that the effect of differential
extinction on completeness is relatively minor, typically $<10\%$ and
at most $20\%$.

\subsection{Wide field} 

\begin{figure}[!htb]
\includegraphics[width=\columnwidth]{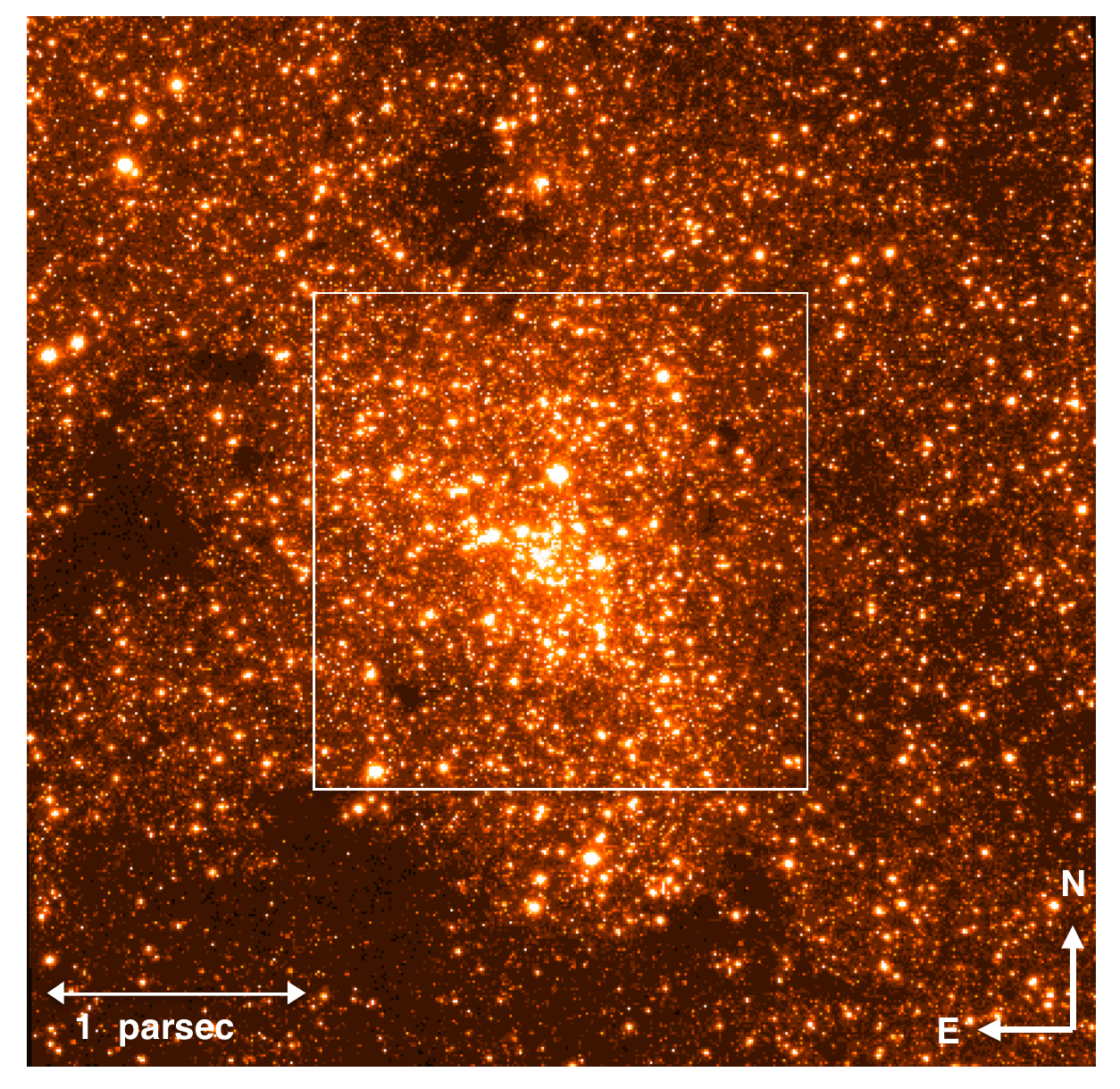}
\caption{\label{Fig:widefield} Wide-field mosaic the observations from
  11th May 2011. The field-of-view is $1.5'\times1.5'$. The field of
  about $40"\times40"$ that corresponds to the deep imaging data is
  marked by a white square. }
\end{figure}

The 2011 data are of excellent quality, but relatively shallow. On the
other hand, there are $16$ pointings that cover a field of about
$1.5'\times1.5'$, compared to the smaller fields of about
$40"\times40"$ covered by the other NACO observations. The 2011
observations are therefore ideally suited to extend the sensitive
central observations with high angular resolution, albeit somewhat
shallower, data out to larger distances.
Fig.\,\ref{Fig:widefield}. Although the wide field data are
  shallow, in the sense of small total exposure time, they are used
  here at large projected radii, $R$. Since crowding, not integration
  time, is the main factor that limits source detection at the GC, the
  50\% completeness limit of the wide field data is still as low as $K_{s}\approx18.5$ at
  $R>0.5$\,pc. They are therefore ideal to be combined with our deep
  data at smaller $R$. 

\begin{figure*}[!htb]
\includegraphics[width=\textwidth]{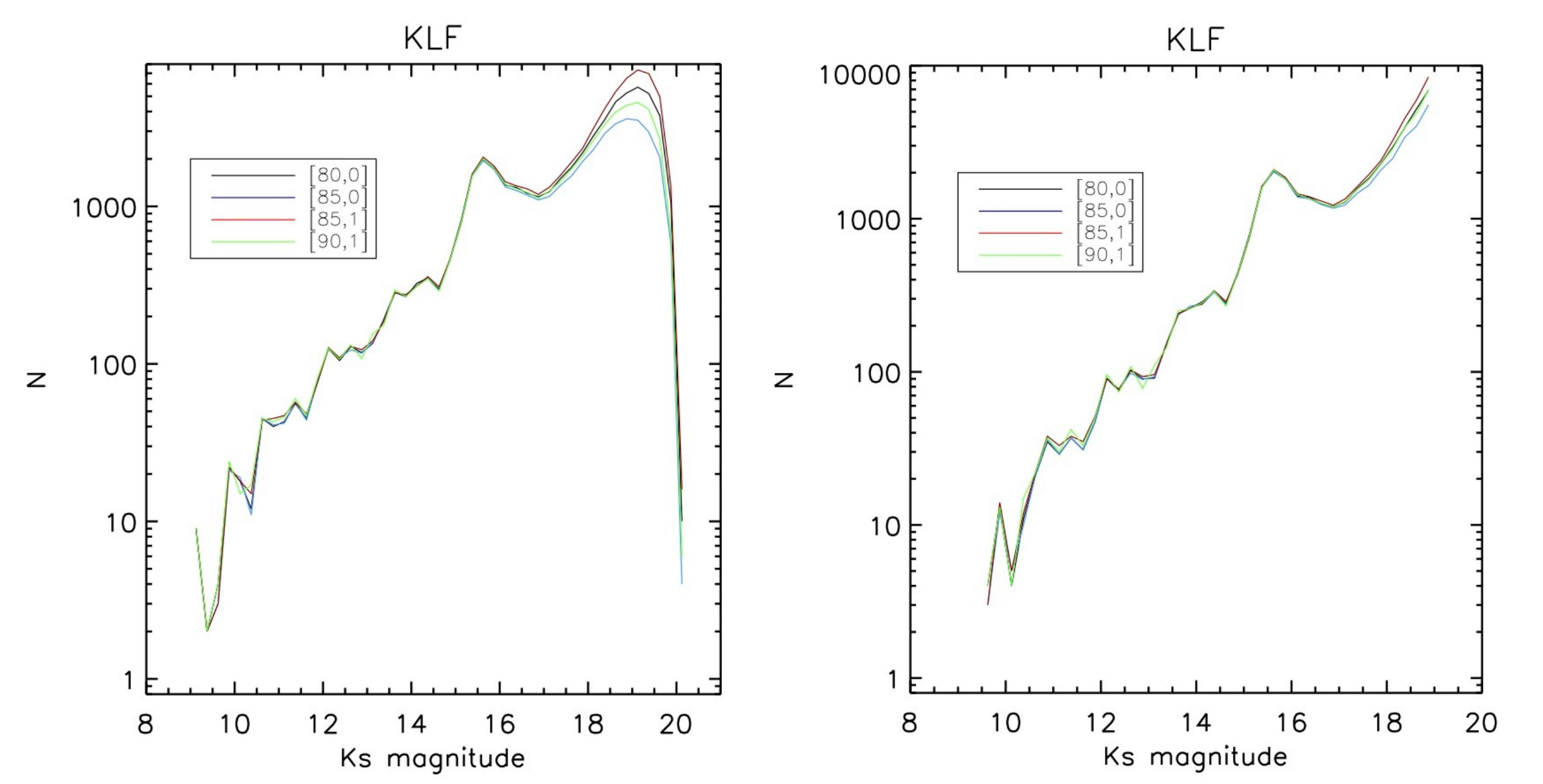}
\caption{\label{Fig:KLF}   KLF
  for the deep $K_{s}$ mosaic. The different colours correspond to the
  different combinations of the values of $min\_correlation$ and
  $deblend$, as listed in the legend (see also
  section\,\ref{sec:detection}). Left: Raw KLF. Right: Completeness-
  and extinction-corrected KLF.}
\end{figure*}

In order to deal with the distortions of the NACO S27
camera \citep[see, e.g.][]{Trippe:2008it,Schodel:2009zr} we aligned
each pointing of the NACO mosaic with a reference frame created from
positions measured in HST WFC3 observations of the same field (Dong et
al., in prep.). We apply variable PSF fitting as we explain in section
$2.4$. In this case,  {\it min\_correlation$=0.8$} and {\it deblend$=0$}
were selected for the {\it StarFinder} parameters.

%--------------------------------------------------------------------
\section{$K_{s}$-luminosity function \label{sec:KLF}}
The goal of this study is to investigate the existence of a stellar
cusp at the GC. This requires us to select stars old enough to have
undergone dynamical relaxation. The relaxation time at the GC is
roughly a few Gyr
\citep[e.g.][]{Alexander:2005fk,Alexander:2011dq}. We specifically
exclude all spectroscopically identified early-type, i.e.\ young and
massive, stars from our sample \citep[using the data
of][]{Do:2013fk}. Unfortunately, spectroscopic stellar classification
is limited to stars of about $K_{s}\leq15.5$ at the GC. For fainter
stars, we can only use their luminosity as a proxy for their
type. Fig.\,16 of \citet{Schodel:2007tw} illustrates the LF, mean
masses, and old star fractions for stars of different magnitudes at
the GC, assuming  continuous star formation at a constant rate
  over the last 10\,Gyr. It shows that we can probe old
($\gtrsim1$\,Gyr), low-mass stars in the range
$15\lesssim K_{s} \leq16$. This is the RC, which dominates all
previous star density measurements. The fraction of old stars rises
again above $\sim$50\% for stars $K_{s}>17.5$, reaching practically
100\% at $K_{s}\approx18$. Therefore, in this study, we focus on the
magnitude ranges $15\leq K_{s} \leq16$, the RC, and
$17.5\leq K_{s} \leq18.5$, i.e.\ the faintest stars accessible by our
data.

We show the $K_{s}$ luminosity function (KLF) determined from our deep
mosaic in Fig.\,\ref{Fig:KLF}. The KLFs corresponding to the four
different {\it StarFinder} parameter settings are shown. The KLF
derived in this work is about one magnitude deeper compared to
previous work \citep[Fig.\,10 in][]{Schodel:2007tw}. This is a
decisive advantage. When we want to probe the existence of a
dynamically relaxed stellar cusp around Sgr\,A*, we need to focus on
stars that are at least several Gyr old. As shown in the illustrative
Figure\,16 in \citet{Schodel:2007tw}, the only magnitude range where
this was previously possible was around the RC
($15.25\leq K_{s}\leq 16.25$). Now, with the deeper data from our new
analysis, we can probe another, fainter magnitude range with a high
fraction of old stars ($17.5\leq K_{s}\leq 18.5$). Also, the stars in
these two brightness ranges have similar masses, which lets us expect
a similar surface density distribution. We discuss in sections
  $5.2$ and $5.3$ the different populations that we can expect in the
  faintest range of stars, based on the latest determined star
  formation history for the GC, and the possible contamination of our
  star counts by stars too young to be dynamically relaxed.

%--------------------------------------------------------------------
\section{The 2D density of old stars in the GC}

\begin{figure}[!htb]
\includegraphics[width=\columnwidth]{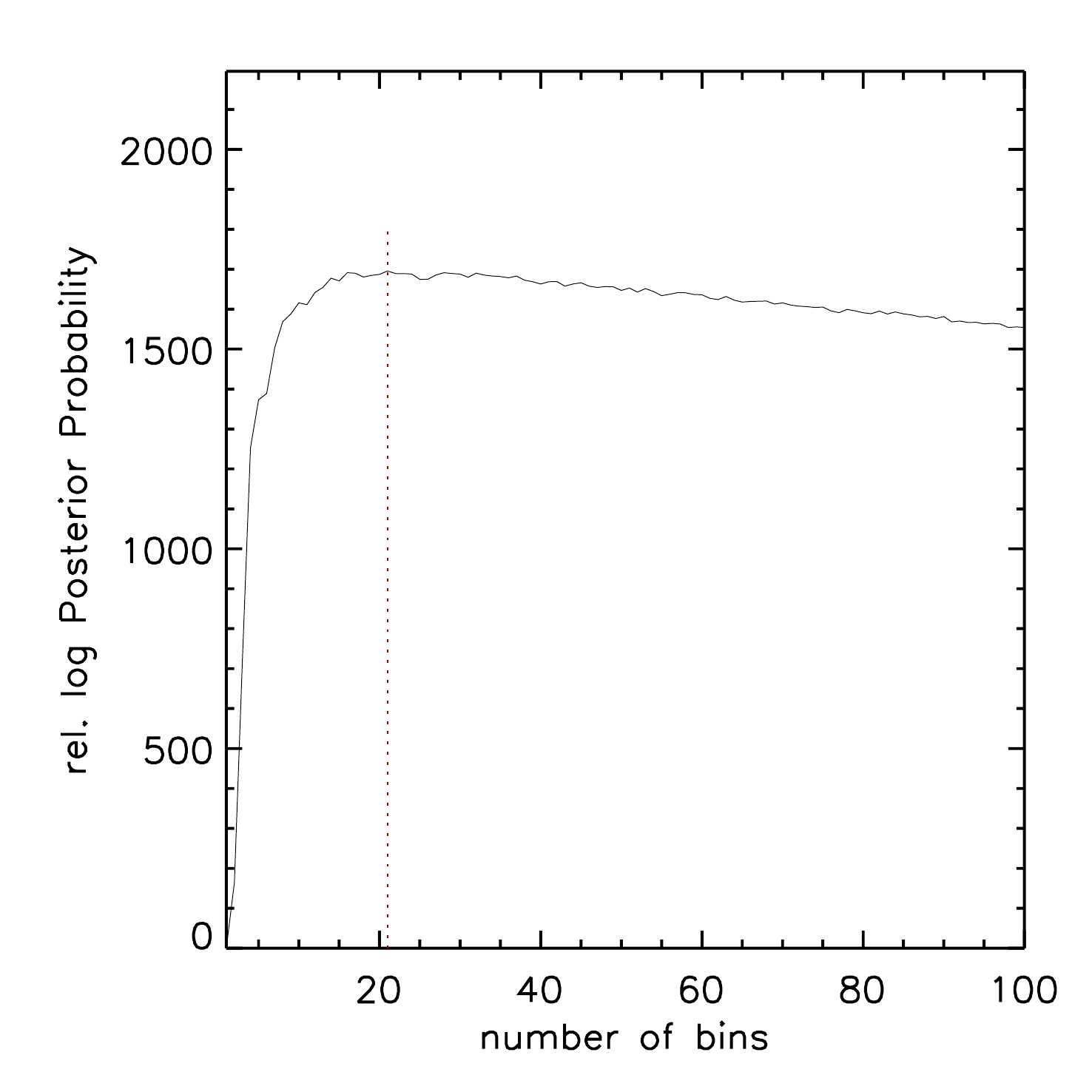}
\caption{\label{Fig:optbin}Optimal binning. We show the RLP
  as a function of the number of bins. The maximum for the
  RLP is reached for 21 bins (red dotted line).}
\end{figure}

In order to analyse the surface density of stars at the Galactic
Centre, we assume that the underlying spatial distribution of the
stars in the central parsec is spherically symmetric. Although the
nuclear cluster is flattened, a spherical approximation should be
acceptable at projected radii $R\lesssim2$\,pc because the difference
between the density profiles along the orthogonal directions of maximum
difference is only on the order of $10\%-20\%$ in this region
\citep[see][]{Schodel:2014fk,Fritz:2016fj}.

\subsection{Star counts in the central parsec}

We computed the azimuthally averaged stellar surface densities in annuli
around Sgr\,A*. It is important to choose a number of bins
sufficiently large to capture the major features in the data while
ignoring fine details due to random fluctuations. Following the
studies of \citet{Knuth:2006pK} and \citet{Witzel:2012fk} we first
determine the best bin size. The dependence of the Relative
Logarithmic Posterior Probability (RLP) on the bin number for pointing
1 is shown in Fig.\,\ref{Fig:optbin}. The maximum for the RLP for the
star number is reached for 21 bins, and the best bin size is $1''$. We
applied this methodology for all pointings, with similar results.

We computed extinction and completeness-corrected stellar surface
densities for the stars detected in the different {\it StarFinder}
runs and for the different pointings (the second step in
  Fig.\,\ref{Fig:Scheme}). At this point we included the uncertainties
of the different correction factors into the uncertainties of the
stellar surface densities. For faint stellar magnitudes, we masked
the regions with completeness below $30\%$. We tested the effect of
different masks, with {\it completeness$<30\%,40\%,50\%$}
respectively, and found that the results did not vary significantly.
Finally, we obtained the surface density profiles for each set of
  {\it Starfinder} parameters by combining the measurements on the
  four pointings in a weighted mean (the third step in
  Fig.\,\ref{Fig:Scheme}).  Finally, we combined the surface density
profiles obtained with the four settings of the {\it StarFinder}
parameters. Mean densities and standard deviations were computed and
all uncertainties were quadratically combined (the last step in
  Fig.\,\ref{Fig:Scheme}).

\begin{figure}[!htb]
\includegraphics[width=\columnwidth]{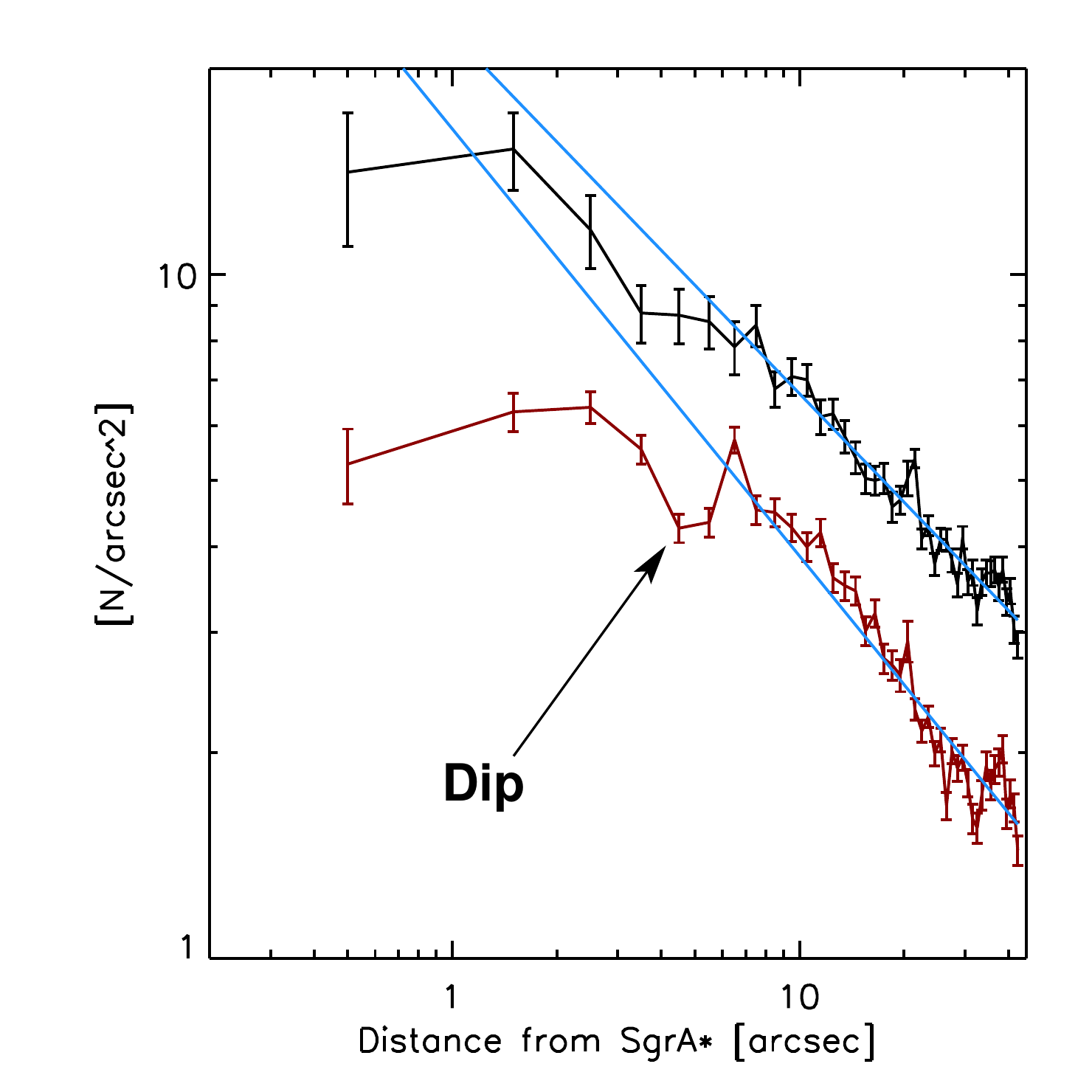}
\caption{\label{Fig:surface_wide} Combined deep field plus wide field
  surface density plots for stars in the magnitude intervals
  $12.5\leq K_{s}\leq16$ and $17.5\leq K_{s}\leq18.5$. The blue lines
  are simple power-law fits to the data at
  $0.2\,\mathrm{pc}\leq R \leq 1.0\,\mathrm{pc}$.. Tables with the
    stellar  surface  density 
data have been made available at the CDS}.
\end{figure}

\subsection{Star counts beyond $20"$}

In order to study the stellar number density in a broader range of
distances from Sgr\,A* we analysed the large mosaic image from the
2011 data. We did not apply any
extinction and completeness corrections because the effect of the
extinction correction is small and crowding does not pose any serious
problem at $R>20"$ and with the high angular resolution data used
here. We did, however mask all the regions occupied by the dark clouds
that can be seen show in Fig.\,\ref{Fig:widefield}. Finally, we
combined the data from the deep field and the wide field. The surface densities from the
wide field data were scaled to the completeness and extinction
corrected ones from the deep images in the overlap region from
$R=10''-20''$.

Figure\,\ref{Fig:surface_wide} shows the combined number density plots
for stars in the magnitude range $12.5\leq K_{s}\leq16.0$ and
$17.5\leq K_{s}\leq18.5$ stars (right), along with simple power-law
fits. The brighter magnitude interval was chosen to trace bright, old
giant stars, based on current estimates of the star formation history
in the central parsec (see section\,\ref{sec:tracers} and
Fig.\,\ref{Fig:old_frac}). The faint magnitude interval corresponds to
the faintest stars that we detect with completeness higher than 50\%
across the field (see sec.\,\ref{sec:completeness}).

\subsection{Projected surface number density \label{projected}}

\begin{table*}[!htb]
\centering
\caption{Values of the power-law index, $\Gamma$, for the extinction and completeness-corrected surface density profiles.}
\label{Tab:Gamma} 
\begin{tabular}{lllll}
\hline
\hline
ID & Fit range (pc) &  Magnitudes range ($K_{s}$) & $\Gamma$ & $\chi^{2}_{reduced}$\\
\hline
1 & 0.04-0.5 & 17.5-18.5 & 0.36$\pm$ 0.04 & 0.8\\
2 & 0.04-0.5 & 12.5-16.0 & 0.24$\pm$ 0.02 & 4.9\\
3 & 0.2-1.0 & 17.5-18.5 & 0.53$\pm$ 0.03 & 2.6\\
4 & 0.2-1.0 & 12.5-16.0 & 0.62$\pm$ 0.02 & 5.1\\
5 & 0.04-1.0 & 17.5-18.5 & 0.47$\pm$ 0.02 & 2.8\\
6 & 0.04-1.0 & 12.5-16.0 & 0.45$\pm$ 0.01 & 9.2\\
7 & 0.5-1.5 & 17.5-18.5 & 0.50$\pm$ 0.03 & 3.2\\
8 & 0.5-1.5 & 12.5-16.0 & 0.73$\pm$ 0.03 & 4.1\\
9 & 0.5-2.0 & 17.5-18.5 & 0.50$\pm$ 0.03 & 3.1\\
10  & 0.5-2.0 & 12.5-16.0 & 0.66$\pm$ 0.03 & 4.6\\
\hline
\end{tabular}
\end{table*}

Simple power laws of the form $\Sigma(R)\propto R^{-\Gamma}$ were fit
to the surface number densities, where $\Sigma$ is the surface number
density, $R$ the projected radius, and $\Gamma$ the power-law
index. We fitted the power laws to the data in different distance
ranges. The corresponding power law indices and the $\chi^{2}$ values
of the fits are listed in Tab.\,\ref{Tab:Gamma} and the data and one
of the fits are shown in Fig.\,\ref{Fig:surface_wide}. All formal uncertainties from the fits were rescaled to a reduced  $\chi^{2}=1$ here and in the rest of the paper. We observe
that: (1) a simple power-law provides a better fit to the faint stars
($K_{s}\approx18$) than to the bright giants. (2) The value of
$\Gamma$ depends on the range in $R$ used for the fitting, with a
tendency for a steeper power-law at greater distances. (3) The data
point at $R<1"$ lies below the fit in all cases. This region is the
most crowded region with a possibly altered stellar population (the
S-stars) and we omitted it therefore from our fits. (4) The giant
stars show a flat, or even decreasing surface density at
$R\lesssim8"$, in agreement with what has been found before
\citep{Buchholz:2009fk,Do:2009tg,Bartko:2010fk}.  They also display a
significant dip around $R=0.2$\,pc ($5''$) in the density profile and,
possibly, an excess at $R\approx7"$, that are also visible in the
works of \citet{Schodel:2007tw} and \citet{Buchholz:2009fk}.

The projected surface number density of the stars in the interval
$17.5\leq K_{s}\leq18.5$ can be described well by a single power-law.
Its mean value and standard deviation, taking into account the
different fitting radii in Tab.\,\ref{Tab:Gamma} is
$\Gamma_{faint}=0.47\pm0.07$ The giants present a somewhat different
picture: A single power-law provides only a good fit to the data at
$R\gtrsim8"/0.24$\,pc. To take this into account, we exclude fit ID\,2
from Tab.\,\ref{Tab:Gamma}  and obtain a mean value of
$\Gamma=0.62\pm0.12$.

%--------------------------------------------------------------------
\section{Discussion}

\subsection{Influence of the correction factors}

\begin{figure}[!htb]
\includegraphics[width=\columnwidth]{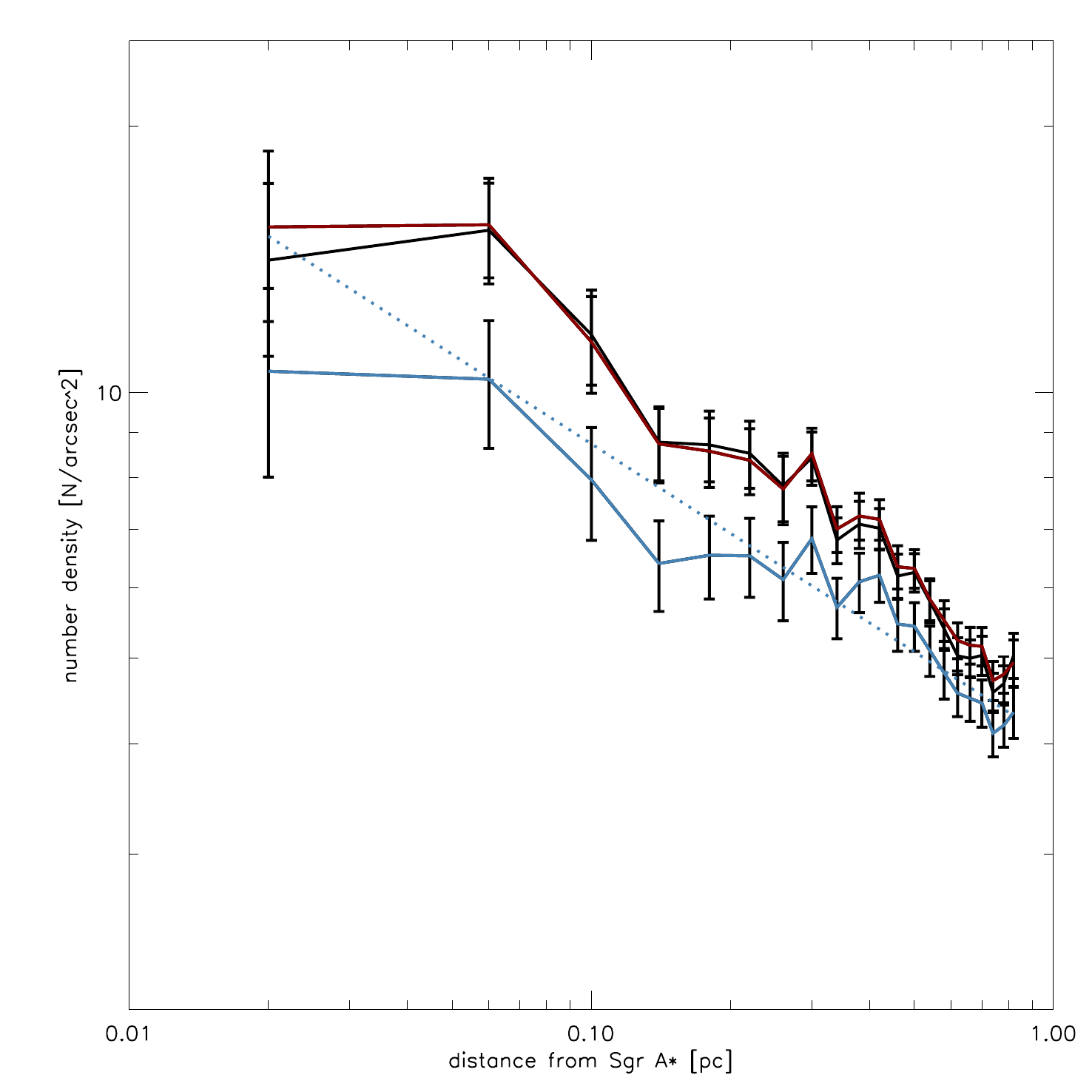}
\caption{\label{Fig:corrections17.5-18.5} Mean surface density profile for stars with ($17.5\leq K_{s}\leq
18.5$), after averaging over the four runs with different {\it
  Starfinder} parameters. Blue: Uncorrected data. Red: Data corrected
for crowding. Black: Data corrected for crowding and extinction.}
\end{figure}

When determining the number density profile, we have to apply
correction factors to compensate the effects of variable stellar
crowding and interstellar extinction.
Fig.\,\ref{Fig:corrections17.5-18.5} shows the measured surface
density profile for stars of magnitude $17.5\leq K_{s}\leq 18.5$ 
  detected in the deep mosaic without any correction, after applying
the completeness correction for crowding, and after applying the
completeness corrections for crowding and extinction. As we can see,
the completeness correction steepens the profile somewhat. The
extinction correction only introduces minor changes because the
azimuthal averaging compensates most of the effects of differential
extinction across the field. Fitting a simple power-law to the
  uncorrected data in the range $0.04\,\mathrm{pc}\leq R\leq1$\,pc, we
  obtain $\Gamma=0.33\pm0.03$ (for the fully corrected data we obtain
  $\Gamma=0.47\pm0.02$, see Table\,\ref{Tab:Gamma}).  The effect of
  crowding correction is almost negligible for the giants brighter
  than $K_{s}\approx16$. We note that the uncertainties of the
  crowding and extinction corrections are included in all error bars
  and will therefore be reflected in the formal uncertainties of the
  best-fit parameters.

  For the wide field data we did not apply any extinction and
  completeness corrections because we lacked the necessary
  complementary wide-field $H-$band imaging data. In any case, the effect of the
  extinction correction on the number density is small and crowding
  does not pose any serious problem beyond $20''$. Hence, the wide
  field data are scaled to the deep data in the overlap region. We
  mask the regions occupied by the dark clouds in the wide field image
  (see Fig.\,\ref{Fig:widefield}) to compute the surface density
  profile. In summary, the applied correction factors, albeit
necessary, do not significantly alter our results. This shows that our
data are robust. We note, however, that the wide field image in
  Fig.\,\ref{Fig:widefield} appears to show that extinction is higher
  at larger $R$, in particular to the west of Sgr\,A*. Since we do not
  correct the magnitudes of the stars detected in the wide image for
  extinction, this may result in giants dropping out of the brightness
  bin considered here. This may explain why the projected surface
  density of the giants appears to show a slightly steeper decrease at
  large $R$ than the surface density of faint stars
  (Fig.\,\ref{Fig:surface_wide}).

\subsection{ Age of tracer populations and possible contamination by
  $\sim$100\,Myr-old stars}\label{sec:tracers}

With an approximate magnitude of $K_{s}=18$, the faintest stars in our
sample are consistent with being (sub-)giants on the ascending branch
or main sequence (MS) stars of $\sim$$2.5$\,M$_{\odot}$.
They could also be pre-MS stars of a few solar masses or less
\citep{Lu:2013fk}.  From what is known about the star formation
history of the NSC we may expect that the majority of stars is old
\citep[$\sim$80\%
of the NSC's mass were formed $>5\,$Gyr
ago, according to][]{Blum:2003fk,Pfuhl:2011uq} and that most of the faint stars in
our sample are thus old, (sub-)giants. 

\begin{figure}[!htb]
\includegraphics[width=\columnwidth]{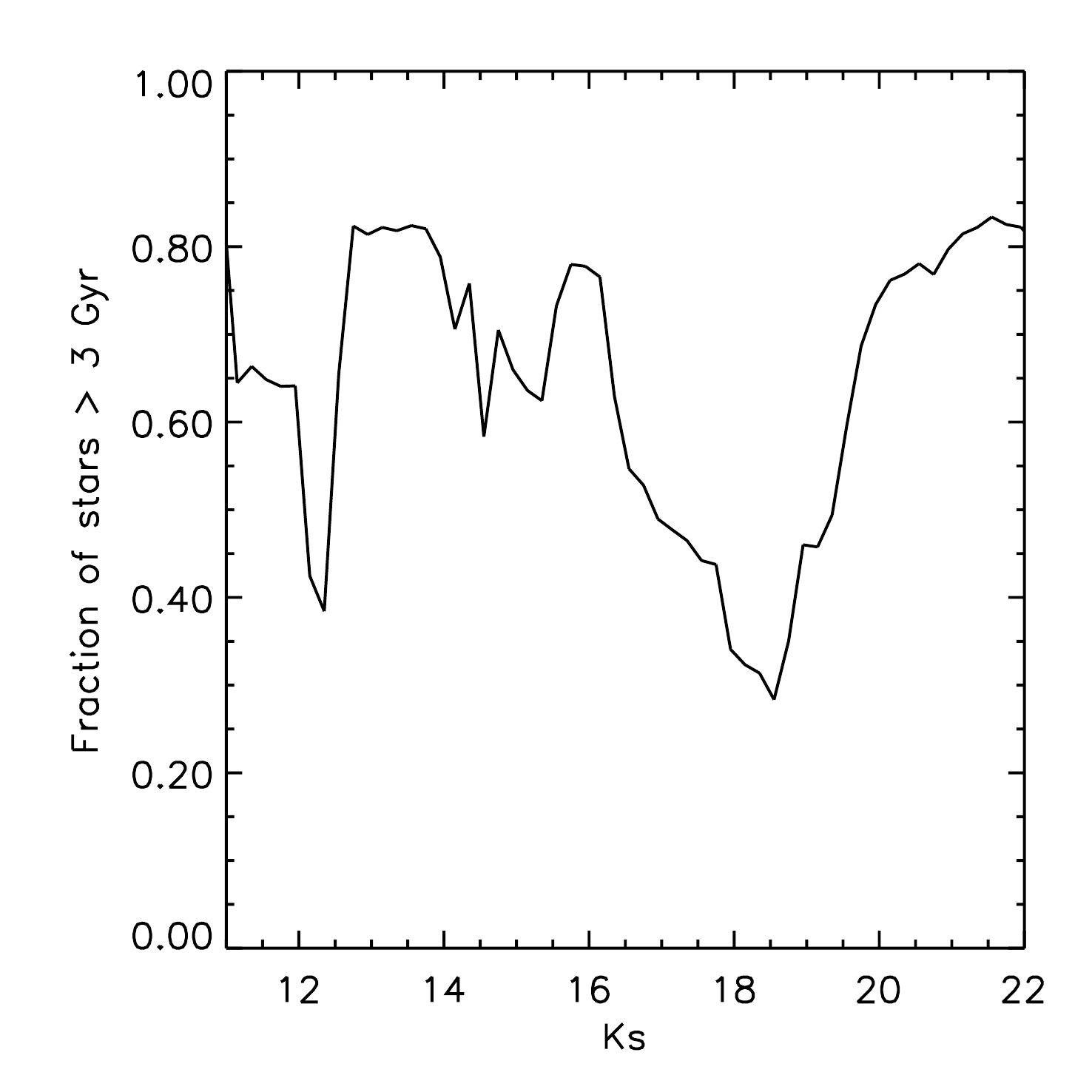}
\caption{\label{Fig:old_frac} Fraction of stars older than 3\,Gyr as a
function of their observed $K_{s}$-magnitude at the GC, computed based
on the SFH derived by \citet{Pfuhl:2011uq}.}
\end{figure}

However, there are two important caveats: (1) We know that a star
formation event created on the order $10^{4}$\,M$_{\odot}$ of young
stars in the region about 0.5\,pc around Sgr\,A*
\citep[]{Bartko:2010fk,Lu:2013fk,Feldmeier-Krause:2015}. Therefore
contamination by pre-MS stars is possible. We will discuss this
possibility in the next subsection. (2) There is evidence that the
star formation rate in the central parsec was high about 100\,Myr ago
\citep[e.g.][]{Blum:2003fk,Nishiyama:2016zr,Pfuhl:2011uq}.

  Using the star formation history (SFH) given by equation (3) in
  \citet{Pfuhl:2011uq}, we calculated the fraction of stars older than
  3\,Gyr as a function of magnitude, as shown in
  Fig.\,\ref{Fig:old_frac}. As we can see, contrary to what we assumed
  in section\,\ref{sec:KLF} based on a model of constant star
  formation rate in the GC \citep{Schodel:2007tw}, the magnitude
  interval around $K_{s}=18$ may be dominated by relatively young,
  dynamically unrelaxed stars. 

  Unfortunately, currently there do not exist any adequate data on the
  age composition of the $K_{s}=18$ stars and on the surface density
  of the potential different populations, which would allow us to
  consider an explicit correction for young stars. Depending on the exact
  properties and spatial distribution of these stars, the cusp
  signature could be enhanced or diminished.  This is a 
  source of systematics of unknown impact and needs to be investigated
  by future research. We note that the spectroscopic classification of
  $K_{s}=18$ stars at the GC is beyond the reach of current
  instrumentation and may require telescopes of the 30m class.

\subsection{Possible contamination by pre-MS stars \label{sec:contamination}}

\begin{table*}[!htb]
\centering
\caption{Parameters used in the estimation of the surface density
  profile of pre-MS stars and resulting corrected $\Gamma$ for the
  density profile of stars with magnitudes  $17.5\leq K_{s}\leq
  18.5$, fitted at $0.04\,\mathrm{pc}\leq R \leq 0.5$\,pc. We test two values of the $\eta$-parameter:  $\eta=1.40$ from
  \citep{Bartko:2010fk} and $\eta=0.93$ from \citep{Do:2013fk}, and
  assume the IMF of \citet{Lu:2013fk}.}
\label{Tab:Par} 
\begin{tabular}{lllll}
\hline
\hline
ID & $\eta^{\mathrm{a}}$ & $\Sigma(2")^{\mathrm{b}}$ & $\Gamma^{\mathrm{c}}$ & $\chi^{2}_{reduced}$\\
\hline
1 & 0.93 & 4.0 & 0.22$\pm$ 0.06 & 0.7\\
2 & 0.93 & 6.0 & 0.13$\pm$ 0.07 & 0.7\\
3 & 1.40 & 4.0 & 0.21$\pm$ 0.05 & 0.7\\
4 & 1.40 & 6.0 & 0.12$\pm$ 0.06 & 0.9\\
\hline
\end{tabular}
\begin{list}{}{}
\item  \textbf{Notes.}
\item[$^{\mathrm{a}}$] Power-law index of the surface-density profile for young stars.
\item[$^{\mathrm{b}}$] Estimated surface density of $K_{s}=18$ pre-MS stars at $R=2"$.
\item[$^{\mathrm{c}}$] Power-law index of the surface-density profile
  of $K_{s}=18$ stars after correction for  pre-MS stars.
\end{list}
 \end{table*}

Our primary goal in this study is to determine the spatial
distribution of the old, relaxed stellar population at the GC. Care
must therefore be taken to exclude young and therefore probably
dynamically unrelaxed stars. We have excluded from the analysis all
spectroscopically identified early-type stars from
\citet{Do:2013fk}. Unfortunately, the limit of spectroscopic
identification of early-type stars with current instruments is
$K_{S}\approx16$ in the GC \citep[with the exception of a few, very
deep exposures of small fields that reached
$K_{s}\approx17.5$, see][]{Pfuhl:2011uq}. As argued in the previous
section, the star formation event that took place a few Myr ago in the
central $R\approx0.5$\,pc means that we have to consider explicitly
the possibility that our stellar surface number densities are
contaminated by pre-MS stars. As we discuss in section $4.3$ in
  Paper II, the possible  contamination of the surface number density
  of $K\approx18$ stars by young stars from the most recent, $\sim$$5$
Myr-old star formation can be relative less important  than the
contamination from other young or intermediate-age populations, with
ages $\leq3$ Gyr (see Figure $7$ in Paper II). While we cannot
constrain the properties of this intermediate-age population due to
lack of data, as explained in the preceding paragraph, we have more
knowledge on the youngest stellar population and can thus consider its
contaminating effect more explicitly.

\begin{figure}[!htb]
\includegraphics[width=\columnwidth]{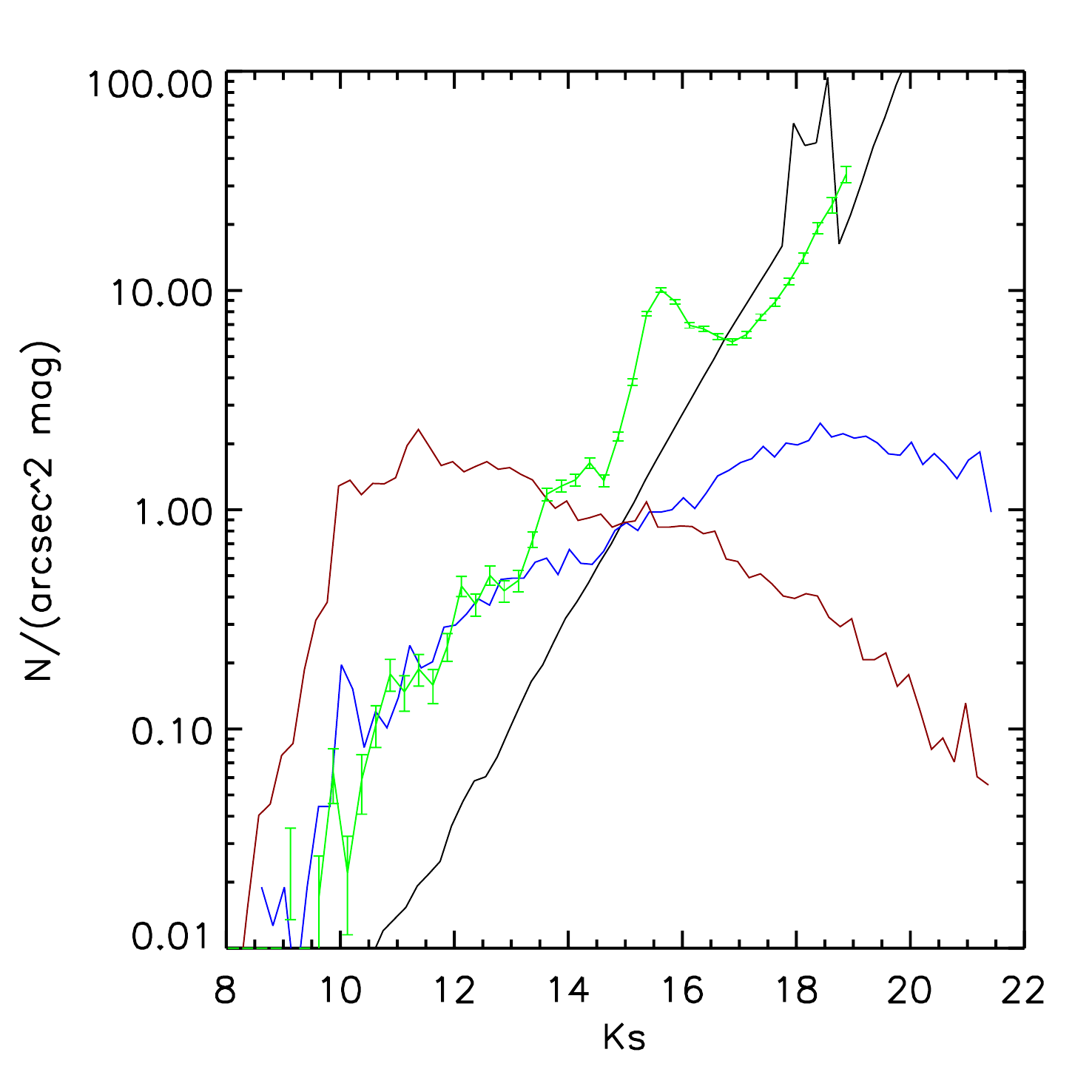}
\caption{\label{Fig:klf_young} Model KLFs for the youngest, $\sim$5\,Myr old
  stellar population at $R=2"$. Black line:
  Kroupa IMF; blue line: IMF from \citet{Lu:2013fk}; red line: IMF
  from \citet{Bartko:2010fk}. The latter two KLFs were created with a
  tool that does not include pre-MS evolution and therefore lack the
  bump at $K\approx18$. The surface density is normalised to $R=2"$
  ($0.08$\,pc), assuming that it rises as $R^{-1}$ towards Sgr\,A*. The
  surface density was normalised with the densities measured by
  \citet{Lu:2013fk}. The green line is the KLF measured from our data
  and normalised to the star counts at $K_{s}=18$. We note that all known
  massive, young stars at $K_{s}\lesssim16$ are excluded from this
  KLF.}
\end{figure}

To estimate the surface density of young stars from the most
  recent star formation event in the GC, we assumed a 5\,Myr old
  cluster of mass $2.5\times10^{4}\,$M$_{\odot}$ of solar
  metallicity. From some experiments with different values we
  concluded that assuming somewhat different ages, masses, or
  metallicities will not change our conclusions significantly. We
  created different present-day model KLFs for this star formation
  event. On the one hand, we used the CMD 3.0 tool
  \citep[http://stev.oapd.inaf.it/cgi-bin/cmd,
  see][]{Bressan:2012xy,Chen:2014nr,Tang:2014rm} with a Kroupa IMF and
  the photometric system based on the works of
  \citet{Maiz-Apellaniz:2006xy} and \citet{Bessell:1990nr}. On the
  other hand, we used the IAC-STAR tool \citep{Aparicio:2004fk} to
  create KLFs with similar parameters, but with a different IMF,
  using, on the one hand, the extremely flat IMF
  $dN/dm \propto m^{-0.45}$ of \citet{Bartko:2010fk} and, on the other
  hand, the steeper, but still top-heavy IMF $dN/dm \propto m^{-1.7}$
  of \citet{Lu:2013fk}. We normalised with the value of $0.3$ stars
  per square arcsec at $K\approx15$ \citep{Do:2013fk,Lu:2013fk} and
  computed the surface density at $R=2"$ assuming that the surface
  density of the young stars follows a power law of the form
  $\Sigma(R)\propto R^{-\eta}$
  \citep[$\eta=0.93-1.1$][]{Bartko:2010fk,Lu:2013fk}.  The resulting
  KLFs for the young stars at $R=2"$ is shown in Fig.\,\ref{Fig:klf_young},
  where we also over-plot the KLF from our full data set, scaling with
  the surface number density of $K_{s}=18$ stars at $R=2"$.

\begin{figure}[!htb]
\includegraphics[width=\columnwidth]{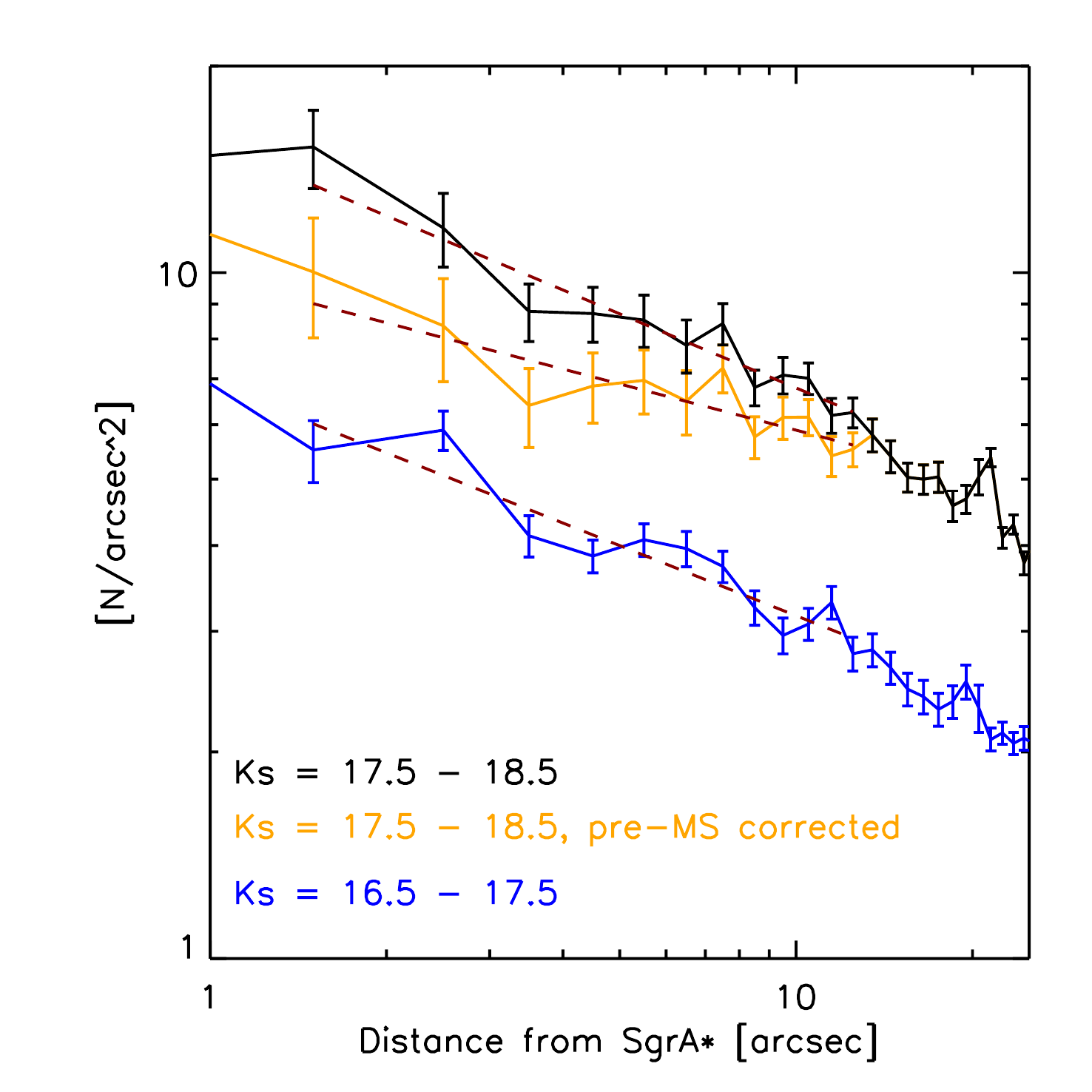}
\caption{\label{Fig:profile_wy} Extinction and crowding-corrected
  surface number density profiles for $16.5\leq K_{s}\leq 17.5$ (blue
  line), $17.5\leq K_{s}\leq 18.5$ (black line), and for
  $17.5\leq K_{s}\leq 18.5$ after correction for potentially present
  pre-MS stars from the most recent star formation event (using
  parameters from ID\,1 in Tab.\,\ref{Tab:Par}). The dashed red lines
  are simple power-law fits in the range
  $0.04\,\mathrm(pc) \leq R \leq 0.5$\,pc. Tables with the stellar
  surface density data have been made available at the CDS.}
\end{figure}

As was pointed out in previous work
\citep[e.g.][]{Paumard:2006xd,Bartko:2010fk,Lu:2013fk}, the IMF of
the most recent star formation event near Sgr\,A* appears to have been
top-heavy. This is supported by our analysis of the KLF here: As can be
seen in Fig.\,\ref{Fig:klf_young} the surface number density of young
stars at $R=2"$ would strongly exceed the measured surface number density of
{\it all} stars in case of a standard Kroupa/Chabrier IMF
\citep{Chabrier:2001yg,Kroupa:2001fv,Kroupa:2003zl}. This problem does
not appear in case of a top-heavy IMF. The IAC-STAR
tool used to infer the top-heavy KLFs does not take pre-MS
evolutionary tracks into account and therefore misses the bump of
stars on the pre-MS at around $K_{s}=18$. The actual surface number
densities can therefore be expected to be a factor of two to three
higher at this magnitude (where we roughly estimated the excess from
the pre-MS onset bump in the KLF that includes pre-MS tracks). If the
IMF of the 5\,Myr-old stellar population is indeed as top-heavy as
suggested by \citet{Bartko:2010fk}, then its contamination of our
number counts can be neglected. For a less extreme IMF, as suggested
by \citet{Lu:2013fk}, the contamination may reach a value up to about
$40\%$ at $R=2"$, but will rapidly diminish due to the steep decrease
of the surface density of the young stars with $R$. 

We have created model surface density distributions for the potential
pre-MS stars, assuming two different parameters for their power-law
index and for their surface density at $R=2"$. Subsequently, those
models were subtracted from the star counts at
$17.5\leq K_{s}\leq 18.5$ and a simple power-law was fitted to the
data at $0.04\,\mathrm(pc)\leq R \leq 0.5$\,pc. The resulting values
of the projected power-law index, $\Gamma$ are listed in
Tab.\,\ref{Tab:Par}. They lie in the range $\Gamma=0.12 - 0.22$,
flatter than for the uncorrected surface density
($\Gamma=0.36\pm0.04$, ID\,1 in Tab.\,\ref{Tab:Gamma}). We note that
this is a conservative scenario, with a high correction factor based
upon the relatively steep IMF of \citet{Lu:2013fk}. When we apply the
extremely top-heavy IMF of \citet{Bartko:2010fk}, we can neglect this
correction for pre-MS stars. This is supported by an additional test,
where we also measured the power-law index of the stars in the
brightness range $16.5\leq K_{s}\leq 17.5$ at
$0.04\,\mathrm(pc)\leq R \leq 0.5$\,pc. It is $\Gamma=0.34\pm0.03$,
consistent with the value for the fainter stars without correction for
pre-MS stars (for the fitting range $0.04\,\mathrm(pc)\leq R \leq 1.0$\,pc it is $\Gamma=0.41\pm0.02$).

The value of $\Gamma=0.36 \pm 0.04$ that we derive for the power-law
index of stars at $K_{s}\approx18$, using only data at a projected
distance of $R\leq0.5\,$pc from Sgr\,A*, lies $9\,\sigma$ away from a
flat core. Even if we take into account the possible
  contamination by pre-MS stars, then we can still exclude a flat
  core, as discussed in the preceding paragraph. We therefore conclude
  that the contamination of the measured surface densities by pre-MS
  stars from the most recent star formation event is probably not
  significant. However, we note that the contamination by slightly
  older stars, from star formation about 100\,Myr ago, is probably a
  more important source of systematic error than the pre-MS stars, as
  discussed in the previous section.

\subsection{3D profile: Nuker fit \label{sec:nuker}}

\begin{table*}[!htb]
\centering
\caption{Best-fit model parameters for Nuker fits to faint stars.}
\label{tab:models}
\begin{tabular}{l l l l l l }
%\noalign{\smallskip}
\hline
\hline
ID  & $r_{b}$ & $\gamma$ & $\beta$ & $\rho(r_{b})$ & $\chi^{2}_{reduced}$\\
& (pc)  & &  & (pc$^{-3}$) \\ 
%\noalign{\smallskip}
\hline
1$^{a}$ & $5.2\pm0.6$ & $1.45\pm0.03$ & $4.6\pm0.7$ & $40\pm7$ & $1.7$\\
2$^{b}$ & $5.0\pm0.5$ & $1.44\pm0.03$ & $4.1\pm0.5$ &  $46\pm7$ & $1.7$ \\
3$^{c}$ & $5.0\pm0.5$ & $1.44\pm0.03$ & $4.1\pm0.5$ & $46\pm7$ & $1.7$\\
4$^{d}$ & $4.8\pm0.5$ & $1.43\pm0.03$ & $3.5\pm0.3$ & $53\pm7$ & $1.7$\\
5$^{e}$ & $4.9\pm0.5$ & $1.42\pm0.03$ & $3.5\pm0.3$ & $53\pm7$ & $1.7$\\
6$^{f}$ & $5.0\pm0.7$ & $1.43\pm0.03$ & $3.6\pm0.6$ & $46\pm14$ & $2.0$\\
7$^{g}$ & $5.0\pm0.5$ & $1.46\pm0.03$ & $3.6\pm0.3$ & $46\pm7$ & $1.7$\\
8$^{h}$ & $5.3\pm0.7$ & $1.41\pm0.03$ & $3.7\pm0.4$ & $39\pm14$ & $1.7$\\
9$^{i}$ & $4.8\pm0.4$ & $1.43\pm0.03$ & $3.5\pm0.3$ & $53\pm7$ & $1.7$\\
10$^{j}$ & $4.3\pm0.5$ & $1.29\pm0.05$ & $3.4\pm0.3$ & $72\pm14$ & $2.0$\\
11$^{k}$ & $4.4\pm0.4$ & $1.29\pm0.05$ & $3.4\pm0.2$ & $72\pm14$ & $1.9$\\
\hline
\end{tabular}
\tablefoot{
\tablefoottext{a}{Fit range: $0.04 \leq R\leq20pc$. Fore-/background emission model 1 of Table\,2 in \citet{Schodel:2014fk}. }\\
\tablefoottext{b}{Fit range: $0.04 \leq R\leq20pc$. Fore-/background emission model 2 of Table\,2 in \citet{Schodel:2014fk}. }\\
\tablefoottext{c}{Fit range: $0.04 \leq R\leq20pc$. Fore-/background emission model 3 of Table\,2 in \citet{Schodel:2014fk}.}\\
\tablefoottext{d}{Fit range: $0.04 \leq R\leq20pc$. Fore-/background emission model 4 of Table\,2 in \citet{Schodel:2014fk}.}\\
\tablefoottext{e}{Fit range: $0.04 \leq R\leq20pc$. Fore-/background emission model 5 of Table\,2 in \citet{Schodel:2014fk}.}\\
\tablefoottext{f}{Fit range: $0.04 \leq R\leq10pc$. Fore-/background emission model 5 of Table\,2 in \citet{Schodel:2014fk}.}\\
\tablefoottext{g}{Fit range: $0.04 \leq R\leq20pc$. Fore-/background emission model 5 of Table\,2 in \citet{Schodel:2014fk}.  Lower integration boundary at $r= R  + 0.01$\,pc}\\
\tablefoottext{i}{Fit range: $0.04 \leq R\leq20pc$. Fore-/background  emission model 5 of Table\,2 in \citet{Schodel:2014fk}. $\alpha=5$.}\\
\tablefoottext{j}{Fit range: $0.04 \leq R\leq20pc$. Fore-/background
  emission model 5 of Table\,2 in \citet{Schodel:2014fk}. Subtracted
  potential contamination by pre-MS stars as in model\,4 of Tab.\,\ref{Tab:Par}.}\\
\tablefoottext{k}{Fit range: $0.04 \leq R\leq20pc$. Fore-/background
  emission model 5 of Table\,2 in \citet{Schodel:2014fk}. Subtracted
  potential contamination by pre-MS stars as in model\,1 of Tab.\,\ref{Tab:Par}.}\\
}
\end{table*}

\begin{figure}[!htb]
\includegraphics[width=\columnwidth]{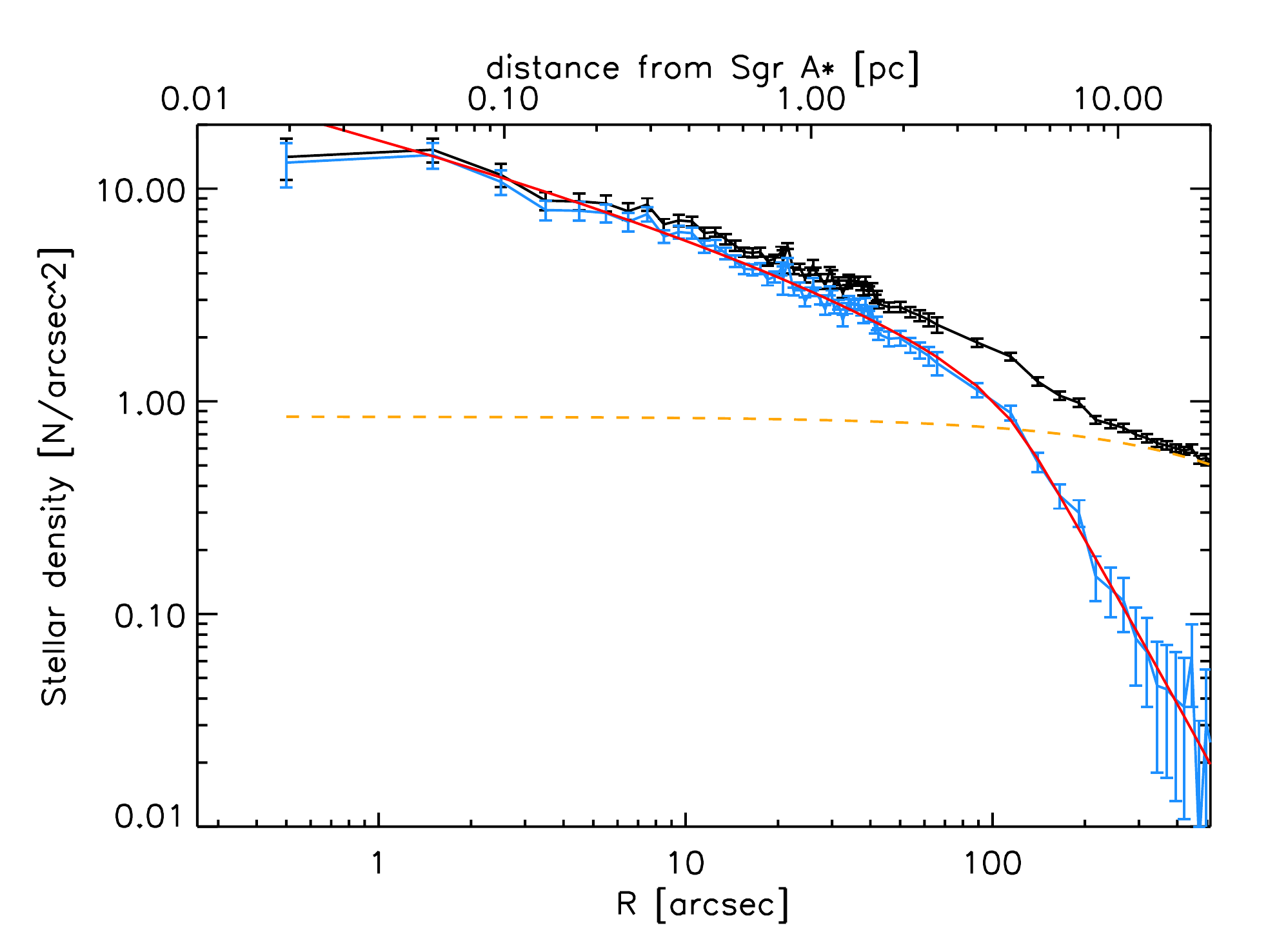}
\caption{\label{fig:nuker} Black: Combined, corrected surface density
  data for stars in the magnitude interval $17.5\leq K_{s}\leq 18.5$ from our deep plus
  wide field image, complemented at large radii by scaled data from
  \citet{Fritz:2016fj}. The dashed orange line is a model for the emission
  from the nuclear disc \citep[model 5 in Table \,2
  of][]{Schodel:2014fk} that is subtracted from the black data points,
  resulting in the blue data points. The red line is a Nuker model fit
  (ID\,5 in Table \,\ref{tab:models}).} 
\end{figure}

As we could see in section\,\ref{projected}, the value of the
projected power-law index depends on the radial fitting range, with a
tendency to steepen at large $R$.  The morphology of the cluster at
large radii will impact the measured projected quantities. For
example, for smaller clusters the inner projected density slope will
appear flatter (see also Fig.\,8 in Paper II). We know that the NSC
shows a steep density decrease at $R\gtrsim2-3\,pc$ \cite[see,
e.g.][]{Launhardt:2002nx,Schodel:2014fk}. Also, we are dealing with a
finite system, with a half-light radius on the order of 5\,pc
\citep{Feldmeier:2014kx,Schodel:2014fk,Fritz:2016fj}. To better
constrain the shape of the NSC, we consider it therefore necessary to
use a 3D model for a fit to the observed projected surface densities.

In order to convert the measured 2D profile into a 3D density law, we
need to deal with projection effects, which requires us to constrain
the surface density on scales larger than what we could measure with
NACO. For this purpose we use the data from \citet{Fritz:2016fj},
which they acquired from observations with NACO/VLT, WFC3/HST, and
VIRCAM/VISTA. To combine these data with ours, we have to assume that,
on large scales, the NSC stellar population is well mixed and that its
average properties (mass function) do not change. We scaled the data
of \citet{Fritz:2016fj} to ours in the range
$0.5\,\mathrm{pc}\leq R \leq 1.0$\,pc. Subsequently, we
  subtracted an estimate of the fore-/background star density using
  models for the non-NSC emission from Table\,2 of
  \citet{Schodel:2014fk}. This simple procedure is possible because
the scale lengths of the latter components are one to several orders
of magnitude larger than the half-light radius of the NSC \citep[see,
e.g.][]{Launhardt:2002nx,Bland-Hawthorn:2016qy}.

We used a 3D {\it Nuker} model \citep{Lauer:1995fk} as given in
equation \,1 of \citet{Fritz:2016fj}:

\begin{equation}
\rho(r) =
\rho(r_{b})2^{(\beta-\gamma)/\alpha}\left(\frac{r}{r_{b}}\right)^{-\gamma}\left[1+\left(\frac{r}{r_{b}}\right)^{\alpha}\right]^{(\gamma-\beta)/\alpha},
\end{equation}

where $r$ is the 3D distance from Sgr\,A*, $r_{b}$ is the break
radius, $\rho$ is the 3D density, $\gamma$ is the exponent of the
inner and $\beta$ the one of the outer power-law, and $\alpha$ defines
the sharpness of the transition.  We fixed the parameter $\alpha=10$,
but explored other values, too (e.g. $\alpha=5$ in fit ID\,9 in
Table\,\ref{tab:models}), with the result that the precise value of
$\alpha$ does not have any significant impact on the best -fit
parameters, in particular on the value of $\gamma$.  We projected the
density onto the sky via an integral as given in equation \,3 in Paper
II and finally we fit the surface density profiles at
$R\leq20$\,pc. As we explain in Paper II, although {\it Nuker}
  model has been previously used for fitting 2D data, we use it as a
  generalisation of a broken power law in order to describe the 3D
  shape of the cluster. In order to determine the fore-/background
  star density we used the S\'ersic models for the non-NSC emission
  listed in Table\,2 of \citet{Schodel:2014fk} and scaled them to the
  data at $R\geq20$\,pc. The results of our fits are listed in
  Table\,\ref{tab:models}.  We also performed a fit with the
  correction for the potential pre-MS stars (ID\,11 in
  Table\,\ref{tab:models}). In Appendix B we explain the computation
  of the systematic uncertainties for the different parameters that
  may result from the deprojection (denoted by the subscript {\it sys}
  in the following). We use the mean of the parameters and their
standard deviation to obtain orientative
values for the average best Nuker model: 
  $r_{b}=4.9 \pm 0.3\pm0.2_{sys}$ pc,
  $\gamma = 1.41\pm0.06\pm0.1_{sys}$, $\beta = 3.7\pm0.4\pm0.1_{sys}$,
  and a density at the break radius of
  $\rho(r_{b})=52\pm12$\,pc$^{-3}$. We note that $\rho(r_{b})$ is
  strongly correlated with the values of the other parameters. Its
  mean value is orientative and we do not cite a systematic error for
  this parameter. The best fit according to model ID\,5 in
  Table\,\ref{tab:models} is shown in Fig.\,\ref{fig:nuker}.

  When we take into account the possible contamination by pre-MS
  stars, then there is a systematic shift towards lower values in the
  best-fit values for $r_{b}, \gamma$, and $\beta$ , as illustrated,
  for example, by fit ID\,11 in Tab.\,\ref{tab:models}: $r_{b}=4.3 \pm 0.5$
  pc, $\gamma = 1.29 \pm 0.05$, $\beta = 3.4 \pm 0.3$.

The Nuker fit shows that the faint stars show a cusp-like distribution
around Sgr\,A*. A flat core can be excluded with high significance. We
explicitly note that in the fits presented in this work we omit the
region $R\leq1"$ ($0.04$\,pc). In this region, the star
counts appear to drop slightly below the expected levels. However, this region
is also the most crowded region, which may lead to strong systematics
in the star counts. Additionally, the stellar population in the
extremely close environment of Sgr\,A* may have been altered, as is
indicated by the presence of the so called 'S-stars', apparently
B-type MS stars that appear concentrated within $R<1"$ of Sgr\,A* and
may have been deposited there by individual scatter or capture events
\citep[see,
e.g.][]{Eisenhauer:2005vl,Genzel:2010fk,Alexander:2011dq}.

\subsection{Distribution of giant stars near Sgr\,A* \label{sec:sfdens}}

\begin{figure}[!htb]
\includegraphics[width=\columnwidth]{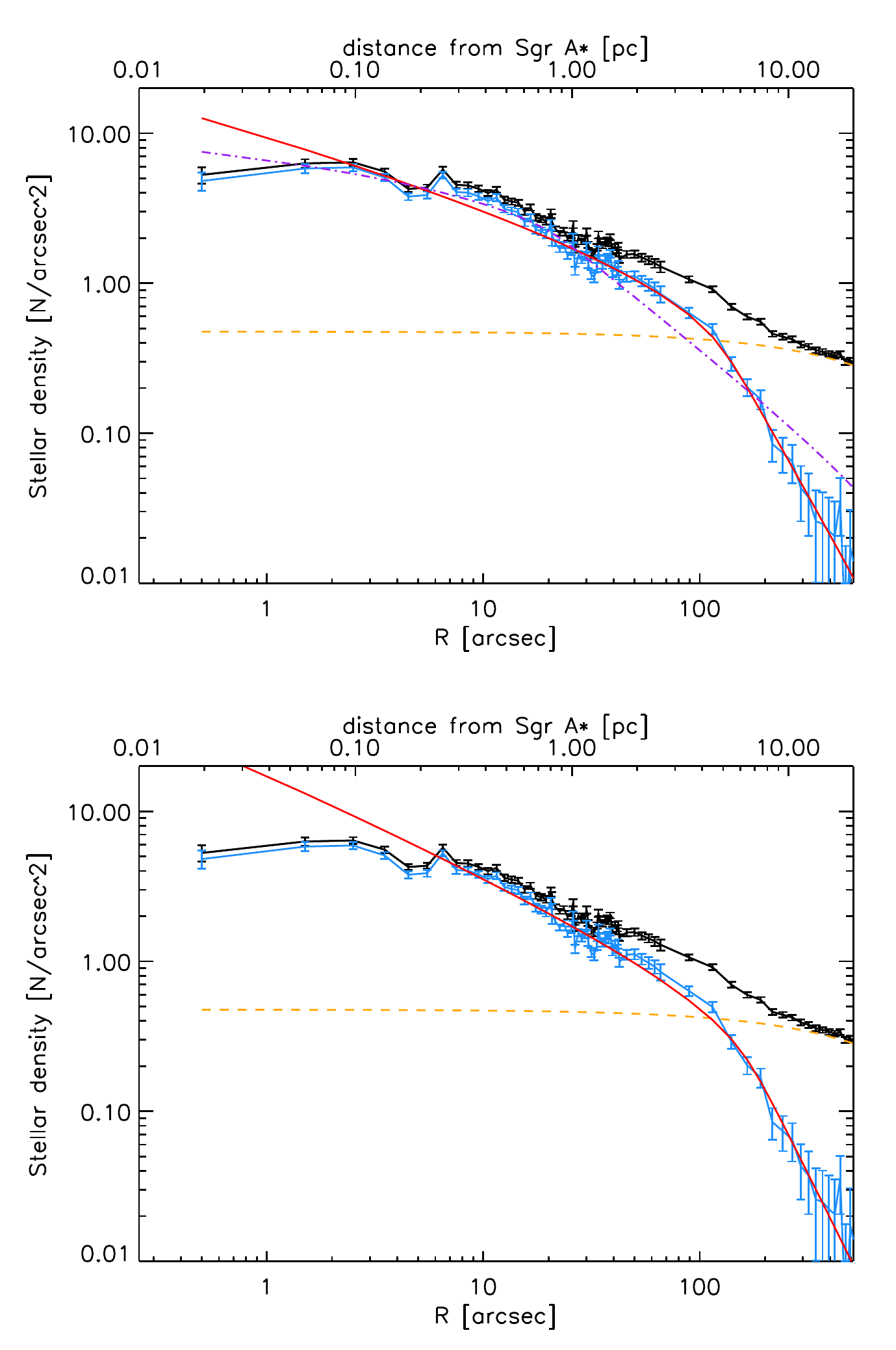}
\caption{\label{fig:nuker_giants} Upper panel: Black: Combined, corrected surface density
  data for stars in the magnitude interval $12.5\leq K_{s}\leq 16$ from our deep plus
  wide field image, complemented at large radii by scaled data from
  \citet{Fritz:2016fj}. The orange line is a model for the emission
  from the nuclear disc \citep[model 5 in Tab.\,2
  of][]{Schodel:2014fk} that is subtracted from the black data points,
  resulting in the blue data points. The red line is a Nuker model fit
  (ID\,1 in Tab.\,\ref{tab:giants}). The dash-dotted purple line is a Nuker
  model fit with $\gamma=1.0$ fixed
  (ID\,7 in Tab.\,\ref{tab:giants}). Lower panel: Like upper panel,
  but Nuker fit from ID\,4 in Tab.\,\ref{tab:giants}.}
\end{figure}

In agreement with previous work, we have found an unexpectedly flat
surface density for RC stars and brighter giants within about
$0.3$\,pc of Sgr\,A*. This indicates a deficit of giants
in this region. Here we produce a 3D Nuker model fit for the giant
stars and try to constrain the number of potentially missing
giants. We proceeded as in section\,\ref{sec:nuker}. At large radii,
we used the data from \citet{Fritz:2016fj} as described in the
preceding section. The resulting best-fit
parameters are given in Table\,\ref{tab:giants}.

\begin{table*}[!htb]
\centering
\caption{Best-fit model parameters for Nuker fits of old giants.}
\label{tab:giants}
\begin{tabular}{l l l l l l }
%\noalign{\smallskip}
\hline
\hline
ID  & $r_{b}$ & $\gamma$ & $\beta$ & $\rho(r_{b})$ & $\chi^{2}_{reduced}$\\
& (pc)  & &  & (pc$^{-3}$) \\ 
%\noalign{\smallskip}
\hline
1$^{a}$ & $5.2\pm0.9$ & $1.45\pm0.04$ & $3.6\pm0.5$ & $23\pm8$ & $5.1$\\
2$^{b}$ & $5.6\pm1.0$ & $1.52\pm0.04$ & $3.7\pm0.6$ & $19\pm7$ & $4.7$\\
3$^{c}$ & $6.3\pm1.2$ & $1.63\pm0.04$ & $3.8\pm0.7$ & $14\pm5$ & $3.8$\\
4$^{d}$ & $6.8\pm1.3$ & $1.70\pm0.05$ & $3.9\pm0.7$ & $11\pm5$ & $3.3$\\
5$^{e}$ & $5.8\pm1.0$ & $1.50\pm0.00$ & $3.7\pm0.6$ & $18\pm6$ & $5.2$\\
6$^{f}$ & $4.5\pm0.9$ & $1.37\pm0.04$ & $3.4\pm0.5$ & $32\pm11$ & $6.4$\\
7$^{g}$ & $0.9\pm0.1$ & $1.00\pm0.00$ & $2.2\pm0.1$ & $572\pm89$ & $7.2$\\
\hline
\end{tabular}
\tablefoot{
\tablefoottext{a}{Fit range: $0.04 \leq R\leq20pc$. Fore-/background emission model 5 of Table\,2 in \citet{Schodel:2014fk}. }\\
\tablefoottext{b}{Fit range: $0.1 \leq R\leq20pc$. Fore-/background emission model 5 of Table\,2 in \citet{Schodel:2014fk}. }\\
\tablefoottext{c}{Fit range: $0.2 \leq R\leq20pc$. Fore-/background emission model 5 of Table\,2 in \citet{Schodel:2014fk}. }\\
\tablefoottext{d}{Fit range: $0.3 \leq R\leq20pc$. Fore-/background emission model 5 of Table\,2 in \citet{Schodel:2014fk}. }\\
\tablefoottext{e}{Fit range: $0.04 \leq R\leq20pc$. $\gamma=1.5$  fixed. Fore-/background emission model 5 of Table\,2 in  \citet{Schodel:2014fk}. }\\
\tablefoottext{f}{Fit range: $0 \leq R\leq20pc$. Fore-/background emission model 5 of Table\,2 in \citet{Schodel:2014fk}.}\\
\tablefoottext{g}{Fit range: $0.0 \leq R\leq20pc$. $\gamma=1.0$  fixed. Fore-/background emission model 5 of Table\,2 in  \citet{Schodel:2014fk}. }\\
\}\\
}
\end{table*}

As can be seen from the reduced $\chi^{2}$ values, the quality of the
fit is significantly worse than for the faint stars, but
improves as we omit the centralmost data points. We use the mean and
error of the mean of the best-fit parameters in Tab.\,\ref{tab:giants}
to obtain orientative values for the average best Nuker model for
giants: $r_{b}=5.7\pm0.8\pm0.2_{sys}$ pc,
$\gamma = 1.53\pm0.13\pm0.1_{sys}$, and
$\beta = 3.7\pm0.2\pm0.1_{sys}$. As we can see, the mean parameters
agree within $1-2\sigma$ with the ones determined from the Nuker fits
to the faint stars, omitting fit 7, which we believe to be not
adequate (see below). The differences between the best-fit values for
giants and faint stars may indicate either systematics that we have
not accounted for or that the two brightness ranges do not trace
populations of similar mean age and therefore dynamical state. As can
be seen in Fig.\,\ref{fig:nuker_giants}, a projected Nuker law can
provide a reasonable fit to the projected surface densities of giants
down to projected distances $R\approx0.1\,$pc. Also, if we consider
the Nuker fits to be reasonable 0th order approximations, then they
are consistent with a cusp-like 3D density distribution of the giants,
in spite of the flat projected density at small $R$. Forcing a
flattish inner cusp, for example by fixing $\gamma=1.0$ as in fit ID\,7 in
Tab.\,\ref{tab:giants} will lead to a bad fit at large distances, as
shown by the dash-dotted purple line in the upper panel of
Fig.\,\ref{fig:nuker_giants}, with parameters that deviate strongly
from the best fit-parameters for all other cases (both star counts in
this work and diffuse light in Paper\,II). We therefore argue that, in
spite of the observed core {\it projection} at small $R$, the observed
surface density of giants in the GC is inconsistent with such a
structure. 

 We point out that this does not contradict previous work. Some
  differences can be explained by the use of different data or
  references to constrain the structure of the cluster on scales out
  to 20\,pc. Also, contrary to other studies, in this work we have
  subtracted the projected density of stars that we do not consider to
  form part of the NSC proper, but to belong to the fore-and
  background population. We do this in order to facilitate comparison
  with theory, which always considers isolated systems. While our
  methodology may make a comparison with other publications therefore
  difficult, we point out that, given the statistical and systematic
  uncertainties, our $\gamma$ for the old stars still overlaps within
  about $2\,\sigma$ with the values given by other work
  \citep[e.g.][]{Do:2009tg,Fritz:2016fj}. Hence, while we find consistently an
  observed {\it projected} flat surface density of giants in the
  innermost few arcseconds, we find that this does not require a flat
  core in 3D.

The observed flat projected profile within $\sim$0.3\,pc and the fact that the
quality of the fit improves when we omit the innermost data points may
indicate that something has altered the apparent distribution of
giants in this region. To estimate the number of potentially
'missing' giants, we focus on the region $R\leq0.3$\,pc. Since it is
impossible to know what would be the 'correct' number density model
for giants stars, we use the following simple approach. We fit
different Nuker laws to the data, where we omit the data inside
$R=0.0,0.1,0.2,0.3$\,pc.  The fit for $R=0.0$ can serve as a benchmark
for the actually measured distribution, while the fits that omit data
approximate the cluster structure without potentially missing stars at
small $R$.  The benchmark fit is shown in the upper panel of
  Fig.\,\ref{fig:nuker_giants}, and the fit omitting data at
  $R\leq0.3$\,pc is shown in the lower panel of the same figure.

Subsequently, we compute the amount of stars at $r < 0.2$\,pc for each
model and compare it to the benchmark solution. The difference in
number provides the estimate of possibly missing stars at
$r < 0.2$\,pc. For the three fits that omit data at small R, this
numbers varies between 40 to 200. While the uncertainty of this crude
estimate is high, it provides us with an idea of the order of
magnitude of the problem. The lesson to take away here is, in our
opinion, that any mechanisms that intends to explain the deficit of
giant stars near Sgr\,A* should be able to be efficient enough to
account for roughly 100 missing giants.

\subsection{Comparison to other work and discussion}

Consistent with previous work, we find indications of a flattening of
the density profile of the RC stars and brighter giants inside
$R\approx8"/0.3$\,pc
\citep{Buchholz:2009fk,Bartko:2010fk,Do:2013fk}. Also, our results
agree well with the fits shown for RC stars in
\citet{Schodel:2007tw} (their Fig.\,17). We also note that the density profile of stars
about a magnitude fainter than the RC was analysed in the latter
work. Although \citet{Schodel:2007tw} used a broken power-law, it is
clear that a single power-law, with $\Gamma\approx0.4$, would also
provide a satisfactory fit to the stars fainter than the RC (see right panel in their
Fig.\,17). \citet{Yusef-Zadeh:2012pd} determined the surface light
density of faint stars in HST/NICMOS images at $1.45, 1.70$, and
$1.90\,\mu$m. After the masking of bright
sources, the stars that dominate the light profiles presented in
\citet{Yusef-Zadeh:2012pd} are probably slightly fainter than the RC
and show a single power-law profile with $\Gamma=0.34\pm0.04$ at
$5"\leq R\leq0.7"$, in agreement with what we find in this work and
with the surface density of stars fainter than the RC shown in \citet{Schodel:2007tw}. To
conclude, our analysis agrees well with previously published results.

There are also several key differences between previous work and our
analysis, which goes beyond the existing state-of-the-art:
\begin{enumerate}
\item We reach about one magnitude deeper in the KLF and are not
  limited/dominated by stars in the RC or bright giants.
\item We focus explicitly on well defined brightness ranges to have a
  constraint on the ages and masses of the stars that we are studying,
  while most previous work generally reported surface density
  profiles for broad ranges of luminosities (but was mostly dominated
  by RC stars).
\item We include high angular resolution data on larger radii, out to
  $R\approx2$\,pc \citep[the only other work to do so was by][]{Fritz:2016fj}.
\item We include the most recent data on the large-scale structure of
  the NSC and thus take into account its global structure. Much previous
  work was focussed on the innermost region near Sgr\,A*, e.g.  $R<20"$
  \citep{Buchholz:2009fk}, $R\leq5"$ \citep{Yusef-Zadeh:2012pd}, or
  $R\leq4"$ \citep{Do:2009tg}. While these studies found
  inconsistencies with the existence of a stellar cusp around Sgr\,A*
  \citep[with the exception of][]{Yusef-Zadeh:2012pd}, they not only
  were dominated by bright stars (see above), but also did not take
  the global picture into account. A stellar cusp is expected to be
  well developed inside the radius of influence, $r_{infl}$, of a
  massive black hole, which is on the order $r_{infl}=3$\,pc for
  Sgr\,A*
  \citep{Alexander:2005fk,Merritt:2010ve,Feldmeier:2014kx,Chatzopoulos:2015yu,Feldmeier-Krause:2017rt}.
\end{enumerate}

The features of the present work have allowed us to infer a more
complete picture of the stellar distribution around Sgr\,A*. In
particular, we could show that the surface density distribution of
stars fainter than the RC are inconsistent with a flat, core-like
distribution with high significance,  even if we take into account
  the possible contamination of the faint star counts by pre-MS stars
  from the last star formation event. An important caveat is, however,
  that the faint stars analysed here may be contaminated considerably
  by dynamically young stars from star formation $\sim$100\,Myr
  ago. Unfortunately, we cannot quantitatively estimate the effect of
  this contamination. As concerns the old, giant stars, we find that
  in spite of their flat projected density at $R\leq0.3$\,pc their 3D
  distribution is consistent with a stellar cusp and definitely
  inconsistent with a flat core.

In Paper\,II we study the diffuse stellar light density -- probably
arising from stars in the magnitude range $K_{s}=19-20$. We find that
the projected diffuse light density can be described well with a
simple power-law with an index $\Gamma_{diffuse}=0.28\pm0.03$, 
  roughly in agreement with, albeit flatter than, the value of
  $\Gamma$ found for the surface number density of faint stars in this
  work. The differences between the surface density profiles of the
  stars in these different magnitude ranges can indicate a different
  age composition (see Fig.\,\ref{Fig:old_frac}). The differences
  between the values of $\Gamma$ may also give us an idea of the real
  uncertainty of our measurements $\Gamma$ , given that both methods
  may be biased by unknown systematic biases. We thus estimate that a
  realistic $1\,\sigma$ uncertainty of $\Gamma$ is on the order
  $\sim$$0.1$.
  For the Nuker fits of the 3D density, we also find roughly similar values
  between the $K_{s}\approx18$
  stellar population and the fainter, unresolved one studied in
  Paper\,II. For the former, we find an inner 3D power-law index of
  $\gamma_{resolved}=1.41\pm0.06\pm0.1_{sys}$,
  for the latter we find $\gamma_{unresolved}=1.13\pm0.02\pm0.05{sys}$.
 This allows us to claim that there exists a stellar cusp around the
massive black hole Sgr\,A*. A flat core can be rejected with very high
confidence.

 The values for the break radius inferred from faint star counts
  and from diffuse light in Paper\,II are different $r_{b,resolved}=5.0\pm0.2\pm0.2_{sys}$
and $r_{b,unresolved}=4.3\pm0.5$, but agree within their uncertainties.
We note that the values for $r_{b}$
derived here are significantly different from what was estimated by
\citet{Fritz:2016fj}. This is because the surface density in the
latter work was dominated by giants/RC stars in the innermost
regions. Also, \citet{Fritz:2016fj} fit the surface density of all
stars in the GC field (including those from the Galactic Bar and
nuclear disc), while we subtracted a model to only take
stars in the NSC into account. We point out that the Nuker laws
presented here are optimised to describe the intrinsic properties of
the nuclear cluster, but not the overall stellar density at the
GC. The NSC lies, of course, embedded in the overall surrounding
structures (e.g. nuclear disc, Galactic Bar and Galactic Bulge). The work of
\citet{Fritz:2016fj} or \citet{Schodel:2014fk} provides data that
include the large scale stellar structures.

%-----------------------------------------------------------------

\section{Conclusions}

 In this work we have revisited the question of the distribution of
 stars around the massive black hole at the centre of the Milky Way
 because previous studies had come up with the unexpected result that
 there did perhaps not exist any stellar cusp. To overcome existing
 limitations, we used improved analysis techniques and deep, stacked
 images. We could push the completeness limit of star counts about one
 magnitude deeper than what had been done before, to
 $K_{s}\approx18$. This has allowed us to study a stellar population
 that consists probably primarily of several Gyr old,
 $1-2$\,M$_{\odot}$ stars. 

 Contrary to the flattening of the density distribution of giant stars
 near Sgr\,A*, a well known observation that we also reproduced here,
 the projected surface density profile of the $K_{s}\approx18$
   and $K_{s}\approx17$ stars can be described very well by a single
 power-law. We estimate a power-law index of
   $\Gamma=0.47\pm0.07$. This value may be up to $\sim0.3$ lower
   if the star counts are heavily contaminated by pre-MS stars from
   the most recent star formation event. Given the probably highly
   top-heavy IMF of this most recent star formation event, for which
   we present some additional evidence in this paper
   (Fig.\,\ref{Fig:klf_young}), a strong contamination seems unlikely,
   however.

 Both the faint resolved stars and the even fainter, unresolved stars
 studied in Paper\,II show consistently projected power-law
   surface densities around Sgr\,A*. We can thus exclude a core-like
 stellar density distribution with high confidence. 
   Unfortunately, our current best knowledge of the star formation
   history within 1\,pc of Sgr\,A* implies that a large fraction of
   the faint stars may be dynamically unrelaxed. Therefore we cannot
   claim unambiguously the existence of a stellar cusp of old,
   dynamically relaxed stars around Sgr\,A*. However, the fact that we
 find power-law cusps for faint stars in two different brightness bins as
 well as for the diffuse emission that arises from low-mass, probably
 old and dynamically relaxed stars (Paper\,II), makes the existence of
a relaxed cusp highly plausible in our opinion.

 We fit 3D Nuker models to the data to estimate the intrinsic
 structure of the Milky Way's NSC. The
 break radius of the Nuker model lies at about 5\,pc, somewhat
   larger than the radius of influence of Sgr\,A*. This agrees with
 theoretical expectations that the cusp is visible well inside the
 break radius \citep[e.g.][]{Alexander:2005fk}.

 The stellar cusp around Sgr\,A*  inferred from the star counts
   may be somewhat flatter than the theoretically expected value of
 $\gamma_{theor}=-1.5$ for the low-mass stellar component in a
 multi-mass cluster \citep[see][and references
 therein]{Alexander:2005fk}. Some disagreement with theory is to be
 expected, however. After all, the Milky Way's NSC is more complex
 than what was considered in existing theoretical work. For example,
 it is embedded in a complex environment, has a complex star formation
 history, and may have suffered events of infall and accretion of
 smaller clusters
 \citep[e.g.][]{Pfuhl:2011uq,Feldmeier:2014kx,Schodel:2014bn}.  New
 N-body simulations, that were undertaken in parallel to this
 observational work (Paper\, III, Baumgardt, Amaro-Seoane \&
 Sch\"odel, submitted to A\&A) and that included the effect of
 repeated star formation events, find flat cusp slopes in agreement
 with our findings.

Finally, the existence of a stellar cusp implies that the giant stars
around Sgr\,A* do indeed display a deficiency in numbers within a
projected radius of a few 0.1\,pc.  We estimate that on the order 100
giants may be 'missing'. This region overlaps with the
region where we find young massive stars that may have formed in a
dense gas disc a few Myr ago. Repeated collisions with proto-stellar
clumps in this disc may have stripped the giants of their envelopes,
rendering them thus unobservable \citep{Amaro-Seoane:2014fk,Kieffer:2016fk}.

\begin{acknowledgements}
  The research leading to these results has received funding from the
  European Research Council under the European Union's Seventh
  Framework Programme (FP7/2007-2013) / ERC grant agreement
  n$^{\circ}$ [614922]. This work is based on observations made with
  ESO Telescopes at the La Silla Paranal Observatory under programmes
  IDs 183.B-0100 and 089.B-0162. We thank T. Fritz for detailed and
  valuable comments. This work has made use of the IAC-STAR Synthetic
  CMD computation code. IAC-STAR is suported and maintained by the
  computer division of the Instituto de Astrofísica de Canarias. PAS
  acknowledges support from the Ram{\'o}n y Cajal Programme of the
  Ministry of Economy, Industry and Competitiveness of Spain. This
  work has been partially supported by the CAS President's International Fellowship Initiative.
\end{acknowledgements}

\bibliography{/Users/rainer/Documents/BibDesk/BibGC}
%\bibliography{/Users/Laly/BibDesk/BibGC}

%-------------------------------------------------------------------

\appendix

\begin{appendices}
\section{Systematic errors of the 2D fit of the Surface
  Density Profile}

In this appendix we examine some of the  potential sources of systematic
errors in the computation of the power law indices for the surface density fits .

\subsection{{\it Starfinder} parameters, extinction, and completeness}

There can be no absolute certainty in the reliability of source
detection. For that reason, we analysed the images with different
values of the {\it StarFinder} parameters, as described in
section\,\ref{sec:detection}. All error bars used in this work include
the uncertainties due to different choices of {\it StarFinder}
parameters, as well as the uncertainties of extinction and
completeness corrections.

\subsection{Binning}

We analyse the results considering different ways of binning the
data. As we see in Section 4.1, we study the maximum for the RLP for
the star number and we obtain the best bin size for our sample. We
also test other values of binning($0.5"$, $1"$, $1.5"$) and we obtain
a value of the systematic error in the index of the power law less
than $0.01$ for RC stars and less than $0.001$ for fainter
stars($17.5\leq K_{s}\leq18.5$ ).  We conclude that binning is not any
significant source of systematic errors in our analysis.

\subsection{Fitting range}

As we can see in Table\,\ref{Tab:Gamma}, the assumed fitting range can
lead to significant variations in the best value of the projected
power-law index $\Gamma$. It is on the order $\Delta\Gamma=0.1$.

\subsection{Correction for young stars}

When we study  the distribution of fainter stars we have to
consider the possibility that our stellar surface number density is
contaminated by pre-MS stars from the latest star formation event, as
discussed in section\,\ref{sec:contamination}. As we see, we model the surface density
profile of the young stars by a simple power-law and compute the
number of young stars considering different 
scenarios. The uncertainty of this correction depends primarily on the
assumed surface density distribution and mass of the young stars and
is on the order of $0.05$. While the precise age of the cluster does
not matter, the assumed IMF is paramount. Here, we assumed the IMF of
\citet{Lu:2013fk} as a conservative case. If the IMF is chosen as top-heavy
as in \citet{Bartko:2010fk}, then the contamination by pre-MS stars can be considered insignificant

\section{Systematic errors of the 3D fit of the Surface
  Density Profile}

In this section we examine several potential sources of systematic
errors in the computation of the power law indices for the fits of the
3D density (see Section 5.4) for the two analysed ranges of old stars: RC and
fainter stars. 

\subsection{Subtracted contribution from the nuclear disc}

When fitting the Nuker profiles we assume different models for the
contribution of stars that do not belong to the nuclear
cluster. Table\,\ref{tab:models} lists the best-fitting parameters
under those different assumptions, which are included in our final
error estimation.

\subsection{Correction for young stars}

When we compare models 5 and 11 in Table\,\ref{tab:models}) we can see
that the contamination by pre-MS stars can change the best-fit value
of $\gamma$ by about 0.1 dex.

\subsection{Uncertainties from deprojection}

As we see in Section 5.4, we are interested in studying the 3D
structure of the old cluster. We need to convert the measured 2D
profile into a 3D density, so we need to consider the source of
uncertainty from deprojection. For this purpose, a 3D cluster was
simulated in order to project it and apply our procedure for density
estimation. We proceed as follows:
\begin{enumerate}
\item Different 3D clusters are simulated, where $1.000.000$ stars are
  distributed following a 3D Nuker model (equation 1). In order to explore
  which parameters in the Nuker fit are more sensitive to the
  variations of the model, we test different cluster: clusters with a
  systematic variation in the break radius($r_{b}=1.6, 5.8, 6.4$),
  clusters with a systematic variation in the exponent of the inner
  power-law ($\gamma=0.68, 1.74, 2.32$) and clusters with a systematic
  variation in the exponent of the outer power-law($\beta=4.8, 6.4$).
\item Extraction of 100 randomised samples from each model with the same star number of our sample.
\item Computation of projected density.
\item Apply the fit to the samples.
\item Comparison the input parameters in the simulation and computed
  parameters in the samples for each of the different models.
\end{enumerate}

\begin{table*}[!htb]
\centering
\caption{3D cluster simulations to analyse the effect of the
  deprojection from 3D cluster into a 2D one. The three first columns
  show the values of the Nuker parameters adopted for the
  simulations. The last six columns show the standard deviation
    and the median of the parameters obtained with our fit for each of the samples of each
  of the different model clusters.}
\label{Tab:NukerModels}
\begin{tabular}{l l l l l l l l l l}
%\noalign{\smallskip}
\hline
ID  & $r_{b}$ & $\gamma$ & $\beta$ & $\sigma(r_{b}-r_{b0})$ &
 { }$median(r_{b}-r_{b0})$ & $\sigma(\gamma-\gamma_{0})$ &
{ }$median(\gamma-\gamma_{0})$ & $\sigma(\beta-\beta_{0})$ & { }$median(\beta-\beta_{0})$\\
%\noalign{\smallskip}
\hline
1 & 3.20 & 1.16 &3.20 & 0.15 &-0.01 & 0.09 & -0.05&0.07& { }0.01\\
2 & 1.60 & 1.16 &3.20 & 0.06 & { }0.02 & 0.08 &-0.01&0.03& { }0.05\\
3 & 4.80 & 1.16 &3.20 & 0.32 & { }0.04 & 0.09 &-0.05&0.14&-0.05\\
4 & 6.40 & 1.16 &3.20 & 0.67 &-0.23 & 0.12 &-0.11&0.36&-0.18\\
5 & 3.20 & 0.58 &3.20 & 0.18 &-0.04 & 0.19 &-0.21&0.07& { }0.02\\
6 & 3.20 & 1.74 &3.20 & 0.19 & { }0.12 & 0.04 & { }0.09&0.08& { }0.03\\
7 & 3.20 & 2.32 &3.20 & 0.37 & { }0.81 & 0.02 & { }0.18&0.10& { }0.09\\
8 & 3.20 & 1.16 &4.80 & 0.06 & { }0.13 & 0.04 & { }0.03&0.07& { }0.25\\
9 & 3.20 & 1.16 &6.40 & 0.04 & { }0.11 &0.03 & { }0.01&0.11& { }0.34\\
\hline
\end{tabular}

\end{table*}

Table\,\ref{Tab:NukerModels} shows the results. We can see that
$\gamma$ is the least sensitive parameter to the break radius
variations or to beta exponent variations of the model
($\sigma(\gamma)<0.1$) . Only in the case of $\gamma = 0.58$ we can
see a large difference between the gamma input parameter and
the recovered ones. In general, the break radius parameter is the
most sensitive to variations of the model. As we can see, if we ignore
the model with $\gamma=0.58$ as an outlier (it is basically excluded
by our data), safe assumptions for the systematic
uncertainties due to deprojection are 
$\Delta\gamma\approx0.1$; for $\Delta\beta\approx0.1$, and $\Delta r_{b}\approx0.2$.

\end{appendices}

\end{document}